\documentclass[aps,prc,preprint,showpacs,superscriptaddress,floatfix,
nofootinbib]{revtex4}
\usepackage[dvips]{graphicx}
\newcommand{\be}{\begin{equation}} \newcommand{\ee}{\end{equation}}
\newcommand{\ba}{\begin{eqnarray}} \newcommand{\ea}{\end{eqnarray}}
   \newcommand{\nq}{{\bf q}}

\newcommand{\wtil}{\tilde{w}}

\newcommand{\ssmdg}{SSM-$\Delta \,$}

\begin{document}

\noindent
\title{Superscaling of non-quasielastic electron-nucleus scattering}

%

\bigskip
\bigskip
\author{C. Maieron}
\affiliation{Dipartimento di Fisica, Universit\`a del Salento and INFN,
Sezione di Lecce, Via Arnesano, 73100 Lecce, ITALY}
\author{J.E. Amaro}
\affiliation{Departamento de F\'{\i}sica  At\'omica, Molecular y Nuclear,
 Universidad de Granada,
  18071 Granada, SPAIN}
\author{M.B. Barbaro}
\affiliation{Dipartimento di Fisica Teorica, Universit\`a di Torino
    and INFN, Sezione di Torino, Via P. Giuria 1, 10125 Torino, ITALY}
\author{J.A. Caballero}
\affiliation{Departamento de F\'\i sica At\'omica, Molecular y Nuclear,
 Universidad de Sevilla, Apdo. 1065, E-41080 Sevilla, SPAIN}
\author{T.W. Donnelly}
\affiliation{Center for Theoretical Physics,
    Laboratory for Nuclear Science and Department of Physics,
    Massachusetts Institute of Technology, Cambridge, MA 02139, USA }
\author{C. F. Williamson}
\affiliation{Center for Theoretical Physics,
    Laboratory for Nuclear Science and Department of Physics,
    Massachusetts Institute of Technology, Cambridge, MA 02139, USA }
%
\begin{abstract}
The present study is focused on the superscaling behavior of
electron-nucleus cross sections in the region lying above the
quasielastic peak, especially the region dominated by
electroexcitation of the $\Delta$. Non-quasielastic cross sections
are obtained from all available high-quality data for $^{12}$C by subtracting
effective quasielastic cross sections based on the superscaling
hypothesis. These residuals are then compared with results obtained
within a scaling-based extension of the relativistic Fermi gas
model, including an investigation of violations of scaling of the
first kind in the region above the quasielastic peak.  A way
potentially to isolate effects related to meson-exchange currents by
subtracting both impulsive quasielastic and impulsive inelastic
contributions from the experimental cross sections
 is also presented.
\end{abstract}

\date \today
\maketitle

\section{Introduction}
\label{sec:intro}

In recent years scaling \cite{West74,Day90} and superscaling
\cite{DS199,DS299} properties of electron-nucleus scattering have
been studied in great detail. A first line of investigation has been
focused on the behavior of experimental data and on the construction
from them of suitable phenomenological models for lepton-nucleus
scattering \cite{DS199,DS299,MDS02,bar03,amaSSM04,ama07}. A second
line, developed in parallel to the first, has instead been focused
on more theoretical analyses; namely, the superscaling properties of
cross sections obtained within specific nuclear models have been
analyzed with the goals of testing the range of validity of the
superscaling hypothesis and of finding and explaining possible
scaling violations
\cite{ama05,jac05,jac06,amaFSI,jac07,MM07,anto04,anto06,anto07,ADP03,ADP04}.
Lepton-nucleus scattering in the region of the $\Delta$ resonance
has been recently studied in \cite{Ivanov:2008ng,Praet:2008yn} and
an extension of the scaling formalism to neutral current neutrino
processes has also been proposed \cite{amaNC,antoNC,MartinezNC}.

The general procedure adopted in scaling analyses consists of
dividing the experimental cross sections or separated response
functions by an appropriate single-nucleon cross section, containing
contributions from protons and neutrons, in order to obtain a
reduced cross section which is then plotted as a function of an
appropriate variable, itself a function of the energy and momentum
transfer.  If the result does not depend on the momentum transfer,
we say that scaling of the 1st kind occurs. If, additionally, the
reduced cross section has no dependence on the nuclear species, one
has scaling of the 2nd kind. The simultaneous occurrence of scaling
of both kinds is called superscaling.

The superscaling properties of electron-nucleus scattering data in
the quasielastic (QE) region have been extensively studied in
\cite{DS199,DS299} and in \cite{MDS02}: scaling of the 1st kind was
found to be reasonably well respected at excitation energies below
the QE peak, whereas scaling of 2nd kind is excellent in the same
region.  At energies above the QE peak both scaling of the 1st and,
to a lesser extent, 2nd kinds were shown to be violated because of
the important contributions introduced by effects beyond the impulse
approximation: inelastic
scattering~\cite{MDS02,bar03,AlvarezRuso:2003gj}, correlations and
meson-exchange currents (MEC) in both the 1p-1h and 2p-2h
sectors~\cite{ADP03,ADP04,albe89,ama01,Amaro:2002mj,Amaro:2003yd,ama09}, which
mostly reside in the transverse channel.

The variety and complexity of contributions that are present above
the QE peak make it difficult to analyze inelastic data directly in
terms of inelastic scaling variables and functions. Any analysis of
this type requires some kind of theoretical assumption which allows
one to focus on a specific kinematic region, having removed
contributions from other processes (to the degree that one can). In
~\cite{amaSSM04}, the scaling analysis of electron scattering data
was extended to the $\Delta$ resonance region. A
non-quasielastic~\footnote{In ~\cite{amaSSM04} this residual was
called the ``Delta'' contribution, assuming the $\Delta$ to be
dominant. To avoid confusion with later discussions where
$\Delta$-dominance is assumed, in the present work we denote the
entire residual after the quasielastic contribution is removed by
``non-QE''.}  (non-QE)  cross section for the excitation region in
which the $\Delta$ plays a major role was obtained by subtracting
QE-equivalent (see below) cross sections from the data and was found
to scale reasonably well up to the peak. Phenomenologically
determined QE and non-QE scaling functions were then used to obtain
predictions for neutrino cross sections at similar
kinematics~\cite{amaSSM04,ama07}. This approach has been referred to
as the SuperScaling Analysis (SuSA).

In this paper one of our goals is to investigate superscaling, and
its violations, in the region above the QE peak, starting from the
idea presented in~\cite{amaSSM04}. To this purpose, in
Sec.~\ref{sec:formalism} we begin by reviewing the basic formalism
for scaling studies in the QE region; specifically, we summarize the
essential features of the so-called SSM-QE model (to be defined in
that section). We continue in that section by also considering the
$\Delta$ region, reviewing and extending the SuSA approach of
\cite{amaSSM04}. All available high-quality data for $^{12}$C are
reconsidered and analyzed by applying a variety of kinematical cuts
to illuminate the origins of the scaling violations that are
observed. We then proceed to a deeper investigation of these
scaling-violating contributions in the region between the QE and
$\Delta$ peaks. In order to do so, in Sec.~\ref{sec:deltamodel} we
present a model for inelastic electron-nucleus scattering within the
impulse approximation based on the same superscaling ideas of
\cite{amaSSM04}, extending an earlier superscaling-based model for
inelastic scattering \cite{bar03} --- this is the so-called SSM-inel
approach and has a variant denoted SSM-$\Delta$ --- see that section
for specific definitions. These models are used in
Sec.~\ref{sec:results} to compute non-QE superscaling functions and
to compare these with the experimental data and with the SuSA fit
for several choices of kinematics. By subtracting theoretical
inelastic cross sections from the experimental data, in
Sec.~\ref{sec:res} we then use this model to isolate the
non-impulsive components of the cross section and analyze their
behavior in terms of 2p-2h MEC contributions obtained in previous
studies. Finally, in Sec.~\ref{sec:conclu} we summarize our study
and draw our conclusions, including some remarks of relevance for
studies of neutrino reactions with nuclei.

\section{Formalism and previous results}
\label{sec:formalism}
\subsection{Scaling in the QE region: the SSM-QE approach}
\label{sec:scalQE}
Here we present a summary of the relevant formalism for
scaling studies in the QE region, focusing on the formulae and
results which will be used in the rest of our study. We denote this the
SuperScaling Model  for the QE  response functions (SSM-QE). Our purpose is
to illustrate how scaling ideas can be used to motivate the
construction of  superscaling-based models for electron-nucleon
cross sections, in the spirit of \cite{amaSSM04,bar03} where more
extensive discussions can be found.

Within the Relativistic Fermi Gas (RFG) model the only parameter
characterizing the nuclear dynamics is the Fermi momentum $k_F$. In
the following we will retain only the lowest orders in an expansion
in the parameter $\eta_F = k_F/m_N$, $m_N$ being the mass of the
nucleon. Within this approximation the RFG longitudinal (L) and
transverse (T) quasielastic response functions, at momentum transfer
$\nq$ and energy transfer $\omega$, can be written as
\be R_{L,T}^{QE}(\kappa,\lambda) = \frac{1}{k_F}f_{RFG}(\psi)
G_{L,T}^{QE}\, , \label{eq:rltqe0} \ee
where the scaling function is given by
\be f_{RFG}(\psi) = f^L_{RFG} (\psi) = f^T_{RFG} =
\frac{3}{4}(1-\psi^2)\theta(1-\psi^2) \label{eq:frfg} \ee and the
scaling variable $\psi$ is 
\be
\psi = \frac{1}{\sqrt{\xi_F}}
\frac{\lambda-\tau}
{\displaystyle{\sqrt{\left(1+\lambda\right) \tau +
\kappa\sqrt{\tau\left(1+\tau\right)}}}}\,,
\label{eq:psiqe}
\ee
with $\xi_F \equiv \sqrt{1+\eta_F^2} -1$. In the formulae above we
have introduced the usual dimensionless variables: $\kappa \equiv q
/2 m_N$, $\lambda \equiv \omega / 2 m_N$ and $\tau \equiv
\kappa^2-\lambda^2$. Retaining terms only up to order $\eta_F$, the
functions $G_{L,T}^{QE}$ are given by \cite{MDS02}
\ba G_L^{QE} &=& \frac{\kappa}{2 \tau}
\left\{Z\left[(1+\tau)W_{2,p}^{QE} - W_{1,p}^{QE}\right] + N
\left[(1+\tau)W_{2,n}^{QE} - W_{1,n}^{QE}\right]\right\}
\label{eq:GLQE}\\
G_T^{QE} &=& \frac{1}{\kappa}\left\{Z W_{1,p}^{QE} + N W_{1,n}^{QE}
\right\}\,, \label{eq:GTQE} \ea
$W_{(1,2),p(n)}^{QE}$ being the single-proton (-neutron)
electromagnetic structure functions, which are given in terms of
electromagnetic form factors by
\ba
W_{1,p(n)}^{QE} &=& \tau G_{M,p(n)}^2(\tau)\\
W_{2,p(n)}^{QE} &=& \frac{1}{1+\tau} \left[G_{E,p(n)}^2(\tau) + \tau
G_{M,p(n)}^2(\tau)\right]\,. \ea
Given the response functions, the QE cross section is then obtained
as \be \frac{d\sigma}{d\epsilon^\prime d\Omega} = \sigma_M\left[v_L
R_L^{QE} + v_T R_T^{QE}\right]\, ,\label{eq:cross} \ee
where $\epsilon'$ is the outgoing electron energy and
$\Omega=(\theta,\phi)$ is the solid angle for the scattering. Here
$\sigma_M$ is the Mott cross section and $v_{L,T}$ are the usual
kinematic factors.

The expressions above suggested, instead of using the RFG scaling function in Eq.~(\ref{eq:frfg})
that one may work backwards to obtain an experimental scaling function
by dividing the QE cross sections by the quantity
\be
S^{QE} = \sigma_M\left[v_L G_L^{QE} + v_T G_T^{QE}\right]
\label{eq:sqe}
\ee
and then, for use in discussions of 2nd-kind scaling, multiplying
the result by $k_F$: \be f^{QE}(\psi,\kappa) = k_F\displaystyle{
\frac{(d\sigma/d\epsilon'd\Omega)^{exp}}{S^{QE}}} \,.
\label{eq:fqenew} \ee Separate L and T scaling functions can
similarly be obtained as
\be
f_{L,T}^{QE}(\psi ,\kappa) = k_F\frac{R_{L,T}^{exp}}{G_{L,T}^{QE}}\,.
\label{eq:fltqe}
\ee

In our previous analyses of the world $(e,e')$ data we have found
that, for large enough momentum transfer ($q>2 k_F$), 1st-kind
scaling works rather well for values of energy transfer $\omega$
below the QE peak value, $\omega_{QE}$. For large values of $\omega$
deviations are observed, coming from contributions beyond QE
scattering, such as inelastic scattering and MEC effects. A separate
analysis of the longitudinal and transverse channels shows that
these deviations mainly occur in the transverse response, while the
experimental longitudinal reduced cross sections scale much better
and up to larger values of $\omega$. This suggests that we can use
the longitudinal QE experimental scaling function obtained in
\cite{DS199,DS299} to define a phenomenological scaling function.

In particular, assuming that (i) indeed there is a universal
superscaling function and that (ii) it can be identified with the
phenomenological function extracted from the analysis of the QE
longitudinal response, we can now work backwards and use this
superscaling hypothesis to predict cross sections. To be more
specific, we define the superscaling model  for the QE  response
functions ({\it i.e.,} what we are calling the SSM-QE approach in
this work). This consists in using Eq.~(\ref{eq:rltqe0}), but with
\be
  f^{SSM-QE}(\psi)  \equiv f^{QE}_L(\psi)\,.
\label{eq:ssm}
\ee
An important step has been taken here: only the longitudinal cross
sections are employed in defining the phenomenological scaling
function. This choice is based on the fact that the transverse cross
sections can have significant non-QE or non-impulsive contributions,
for instance, the former from inelastic excitations of the nucleon
(importantly the $\Delta$) and the latter from 2p-2h MEC --- see the
discussions to follow in the present work. However, in lowest order
these are not very important in the longitudinal cross section, and
thus it provides the only opportunity to isolate the impulsive
contributions to the nuclear response.

\begin{figure}[t]
\includegraphics[scale=0.9,  bb= 130 520 500 750]{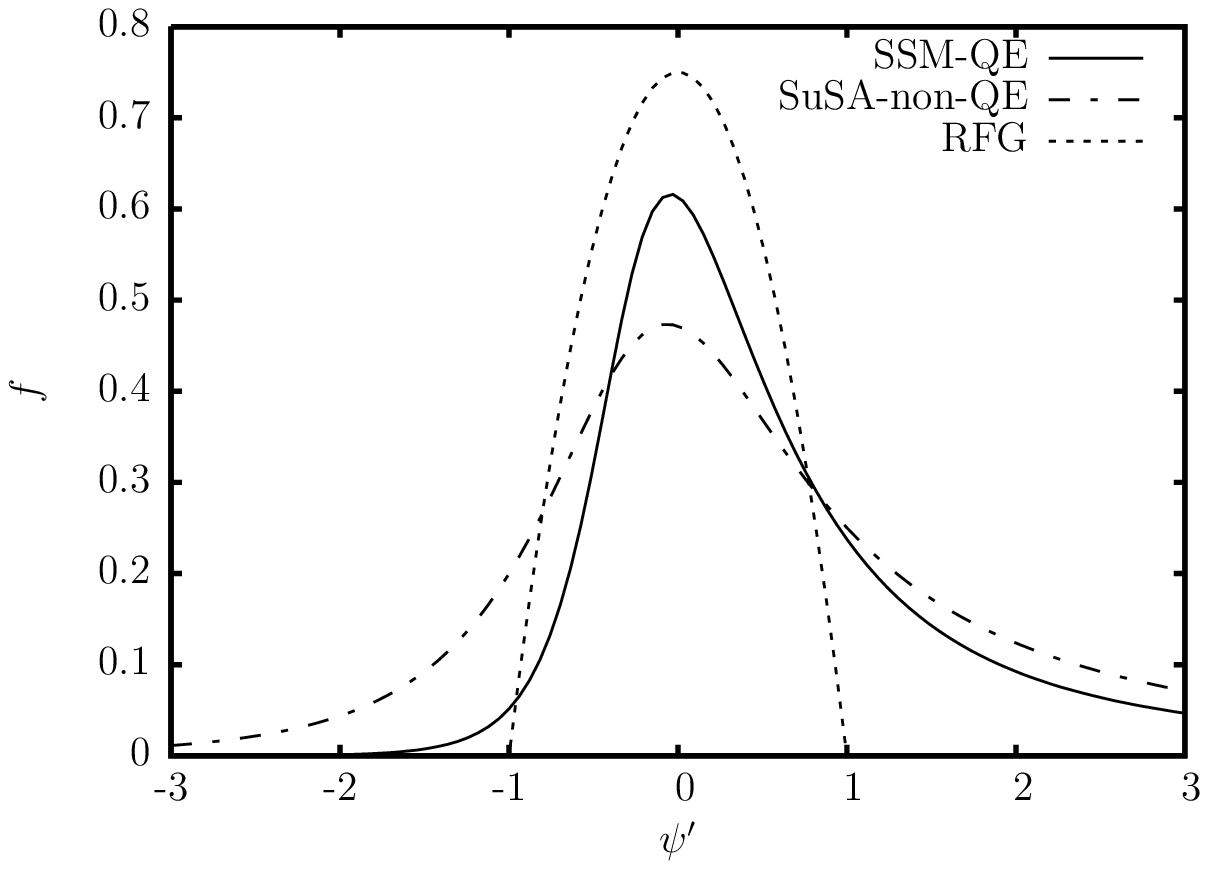}
\caption{Phenomenological fits for the superscaling functions
   $f^{SSM-QE}$ (solid line) and $f^{non-QE}_{SuSA}$
   (dot-dashed) versus the appropriate scaling variable.
The RFG superscaling function is also shown for
  comparison (dotted line).
} \label{fig:phenofits}
\end{figure}

The phenomenological function $f^{SSM-QE}$ employed in the present
approach is shown in Fig.~\ref{fig:phenofits} (solid line), where it
is compared with the RFG scaling function of Eq.~(\ref{eq:frfg})
(dotted line). Also shown is the phenomenological non-QE scaling
function, $f^{non-QE}_{SuSA}$ to be defined below in the following
subsection (dot-dashed line). Focusing on the phenomenological QE
scaling function, one sees that it is significantly different from
the RFG result: it is about 17\% lower at the peak and is
asymmetric, having a tail that extends to higher $\omega$ (in the
positive $\psi^\prime$ direction). In fact, subsequent to obtaining
the phenomenological results shown in the figure~\cite{DS199,DS299},
relativistic mean field theory (RMF) was employed to obtain
theoretical scaling functions. This approach is especially relevant
at high energies where relativistic effects are known to be
important. These RMF studies yielded essentially the same
longitudinal scaling function as the phenomenological
model~\cite{jac06}, and the required asymmetric shape of the scaling
function was obtained theoretically. We shall return below to
comment on the RMF transverse scaling function. Still later a
so-called semi-relativistic approach was pursued~\cite{amaFSI},
again yielding essentially the same results. More recently, a
deceptively simple ``BCS-inspired'' model was developed~\cite{BCS},
with the same outcome: a peak height that is significantly below the
RFG result and an asymmetric shape. Within the flexibility in each
model and the experimental uncertainties one can say that a single
longitudinal QE scaling function has clearly emerged.

In passing, we note that, as is usually done in studies of electron
scattering in order to reproduce the correct position of the QE
peak, in the present study we have introduced a small energy shift
$E_{shift}$. Within the framework of the superscaling formalism
outlined above, this amounts to considering a ``shifted'' scaling \
variable $\psi^\prime$, calculated according to
Eq.~(\ref{eq:psiqe}), but with $\lambda\rightarrow \lambda'=\lambda
- E_{shift}/2 m_N$ and $\tau \rightarrow
\tau'=\kappa^2-\lambda^{\prime 2}$. The values $k_F=228$ MeV/c and
$E_{shift}=20$ MeV have been used in all of the calculations for
$^{12}$C presented here and in the following sections.

Having found that the longitudinal QE scaling function is universal,
whether treated phenomenologically or via models for 1p-1h knockout
reactions, we now discuss the transverse QE response. In most
approaches one finds that once the single-nucleon cross section is
removed in defining scaling functions as above the longitudinal and
transverse answers are basically the same, {\it i.e.,} one has what
has been called scaling of the 0th kind with $f_T (\psi')=f_L
(\psi')$. However, in what is likely the best model employed so far,
the RMF approach cited above, one finds that 0th-kind scaling is
mildly broken for momentum transfers in the 1 GeV region with $f_T
(\psi')>f_L (\psi')$. For instance, at $q=500$ MeV/c (1000 MeV/c)
the transverse RMF scaling function is 13\% (20\%) larger at its
peak than is the longitudinal one. On the other hand, from analyses
of 1p-1h MEC contributions \cite{albe89,ama01,
Amaro:2002mj,Amaro:2003yd,ama09} one sees the opposite behavior,
namely, the 1-body (impulse approximation) and 2-body MEC
contributions to the 1p-1h response, which must occur coherently and
hence can interfere, in fact do so destructively and therefore a
somewhat lower result is found for the total transverse scaling
function. Neither of these effects is seen in the longitudinal
response in leading order. Unfortunately, no single model exists
where one has adequate relativistic content (as in the case of the
RMF approach) and has a consistent way to obtain the MEC
contributions; indeed, the MEC studies cited above could not be
attempted on the same footing as the 1-body RMF computations and
could only be undertaken using much simpler dynamics.

Accordingly, we have no better option at present than to adopt some
working procedure. Henceforth we shall assume that 0th-kind scaling
is obeyed and thus take $f_T (\psi')=f_L (\psi')$ for the
quasielastic response. One should remember, however, that this may
not be completely true and that the QE transverse response could be
either a bit larger or a bit smaller than the one obtained under
this assumption. In Sec.~\ref{sec:res}, where the scaling-based
cross sections are compared with data, we shall return to discuss
these issues in somewhat more detail.

%
\subsection{The SuSA approach to scaling in the $\Delta$ region}
\label{sec:scaldel}

We begin by summarizing the essentials of the SuSA approach taken in~\cite{amaSSM04}, where
non-QE cross sections were obtained from experimental
inclusive inelastic electron-nucleus cross sections by subtracting
QE cross sections given by the SSM-QE procedure described above.
Namely the following cross sections
\be \left( \displaystyle{ \frac{d\sigma}{d\epsilon'd\Omega} }
\right)^{non-QE} \equiv \left( \displaystyle{
\frac{d\sigma}{d\epsilon'd\Omega} } \right)^{exp} - \left(
\displaystyle{ \frac{d\sigma}{d\epsilon'd\Omega} } \right)^{SSM-QE}
\label{eq:sigdelequiv} \ee were obtained as a first step. In the
earlier work it was assumed that $\Delta$-dominance could be
invoked. Namely, in analogy with the QE results of previous section,
a model in which only impulsive contributions proceeding via
excitation of an on-shell $\Delta$ was employed. In that model the
leading-order RFG expressions for the electromagnetic response
function  can be written as~\cite{ama99,amaSSM04}:
\be
  R_{L,T}^{\Delta}(\kappa,\lambda)
  =\frac{1}{k_F}f^\Delta(\psi_\Delta) G_{L,T}^{\Delta}
  \label{eq:rltdel0}
\ee
with $f^\Delta(\psi_\Delta)= f_{RFG}(\psi_\Delta)$ and
\be
  \psi_\Delta = \frac{1}{\sqrt{\xi_F}}
  \frac{\lambda-\tau\rho_\Delta}
  {\displaystyle{\sqrt{\left(1+\lambda\rho_\Delta\right) \tau +
  \kappa\sqrt{\tau\left(1+\tau\rho_\Delta^2\right)}}}}\,,
  \label{eq:psidelta}
\ee
with
\be
  \rho_\Delta = 1 + \frac{\mu_\Delta ^2-4\tau}{4 \tau}\,\,;\,\,
  \mu_\Delta = \frac{m_\Delta}{m_N}
  \label{eq:rhodel}
\ee
and with
\ba
  G_{L}^{\Delta } &=&  \frac{\kappa}{4 \tau} A
  \left[\left(1 + \tau\rho^2_\Delta+1\right)w_2^\Delta -w_1^\Delta\right]
  \label{eq:gldel}\\
  G_{T}^{\Delta } &=&  \frac{1}{2\kappa} A w_1^{\Delta}\,.
\label{eq:gtdel}
\ea
In Eqs.~(\ref{eq:gldel},\ref{eq:gtdel})  the single-hadron
$N\rightarrow \Delta$ structure functions are \footnote{Equations
(\ref{eq:w1del},\ref{eq:w2del}) should be taken
  with $A=Z$ and the $p\rightarrow \Delta^+$ structure functions and with
  $A=N$ and the $n\rightarrow \Delta^0$ structure functions, and then
summed, but since these processes are purely isovector we use $A=N+Z$
  with one choice for the structure functions.}
\ba
  w_1^{\Delta} &=& \frac{1}{2}\left(\mu_\Delta+1\right)^2
  \left(2\tau\rho_\Delta + 1-\mu_\Delta \right)
  \left(G^2_{M,p} + 3 G^2_{E,n}\right)
\label{eq:w1del}\\
  w_2^{\Delta} &=& \left(\mu_\Delta+1\right)^2
  \frac{\displaystyle{\left(2\tau\rho_\Delta + 1-\mu_\Delta \right)}}
  {\displaystyle{1+\tau \rho_\Delta}} \left(G^2_{M,p} + 3 G^2_{E,n} +
  4 \frac{\tau}{\mu^2_\Delta}G^2_{C,\Delta}\right)\,,
\label{eq:w2del}
\ea
where the magnetic, electric and Coulomb form factors are taken to
be
\ba
G_{M,p} &=& 2.97 g_\Delta(\tau)\\
G_{E,n} &=& -0.03 g_\Delta(\tau)\\
G_{C,\Delta} &=& -0.15 G_{M,p}(\tau)\,,
\ea
with
\be
g_\Delta(\tau) =
\frac{1}{\sqrt{1+\tau}}\frac{1}{\left(1+4.97\tau\right)^2}\,.
\ee

Starting from these expressions and assuming that the only non-QE contributions
arise from this $\Delta$-dominance model
one can define a superscaling function in the region
of the $\Delta$ peak as follows
\be
  f^{non-QE}(\psi_\Delta) \equiv k_F
  \frac{\displaystyle{\left( \frac{d\sigma}{d\epsilon'd\Omega} \right)^{non-QE}}}
       {S^\Delta}
\label{eq:fdelta}
\ee
with
\be
  S^{\Delta }\equiv \sigma _{M}
  \left[ v_{L}G_{L}^{\Delta }+v_{T}G_{T}^{\Delta}
  \right] .
\label{eq:sdelta}
\ee

\begin{figure}[th]
\includegraphics[scale=0.75,  bb= 100 300 500 750]{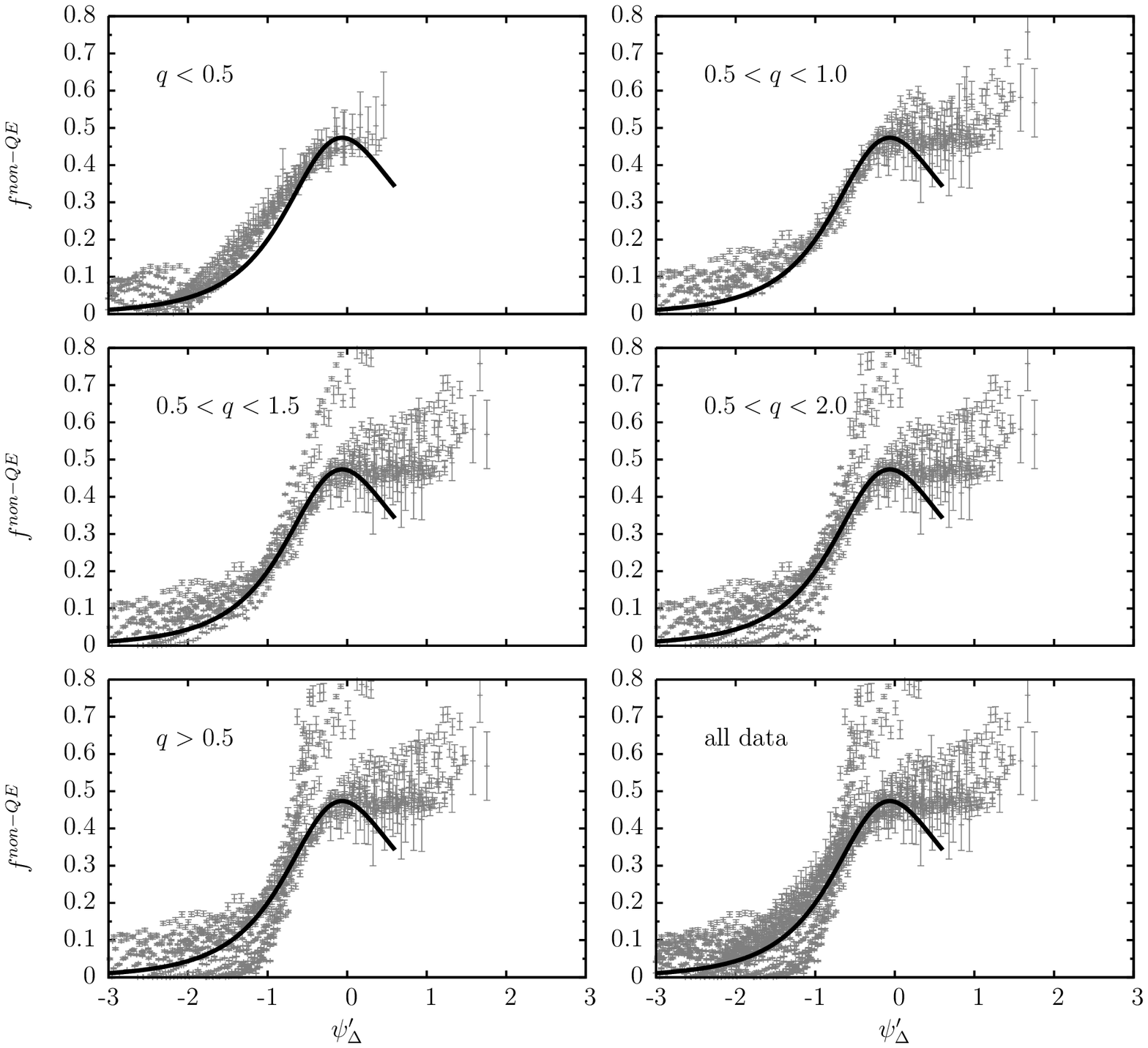}
\caption{``Experimental'' superscaling
  function $f^{non-QE}$ for $^{12}$C, obtained by applying
the QE-subtraction procedure described in the text to the available
experimental data for $^{12}$C. The function is plotted versus the
$\Delta$ scaling variable $\psi_\Delta'$. Kinematical cuts on the
values of the momentum transfer $q$ (in GeV/c) are considered, as
indicated in each panel. A phenomenological fit of the non-QE
superscaling function, $f^{non-QE}_{SuSA}$, is also shown for comparison by the solid
line. } \label{fig:fdelnew}
\end{figure}

We have performed an analysis similar to that presented
in~\cite{amaSSM04} and, focusing on scaling of the 1st kind, we have
considered all available high-quality data of inelastic electron
scattering cross sections on
$^{12}$C~\cite{Whitney:1974hr,Barreau:1983,
Sealock:1989,Day:1993,O'Connell:1987ag,Baran:1988,
Arrington:1995hs,Arrington:1998,Benhar:2006wy,archive}. The
functions $f^{non-QE}$ we obtain are shown in
Fig.~\ref{fig:fdelnew}. Note that, as above, we have introduced a
small energy shift $E_{shift}$. In employing
Eqs.~(\ref{eq:psidelta}) and (\ref{eq:rhodel}) we do as in the QE
case and replace $\lambda$ by $\lambda'$, and $\tau$ by $\tau'$. As
before, for $^{12}$C the values $k_F=228$ MeV/c and $E_{shift}=20$
MeV have been used in all of the calculations presented here and
below.

Overall we see a tendency for coalescence below  and up to the $\Delta$
peak for some, but not all, of the data. Specifically, for kinematics lying below the $\Delta$ peak
($\psi^\prime_\Delta = 0$) these non-QE results scale reasonably
well given the assumption of $\Delta$-dominance, showing scaling
violations at the level of roughly 0.1 units of scaling function,
versus the QE peak value of about 0.6, namely scaling violations of
approximately 15--20\%. As discussed in more detail later, since we
cannot have any inelasticity over much of this kinematic range
(being below pion production threshold) one must suspect that
effects such as from 2p-2h MEC contributions are playing a
non-trivial role. Nevertheless, accepting this as a measure of the potential
uncertainty in following the straightforward SuSA approach,
in~\cite{amaSSM04} an empirical fit to these results, $f^{non-QE}_{SuSA}$, was obtained
and then used to predict neutrino-nucleus cross sections in the
$\Delta$ region. It should be stressed that the assumption of $\Delta$-dominance is clearly
only an approximation; in the following sections we present a more microscopic approach
in which the superscaling approach discussed above for the QE region is extended to the inelastic region
(denoted the SSM-inel approach; see Sec.~\ref{sec:deltamodel}) and where non-impulsive 2p-2h MEC effects are considered
separately (see Sec.~\ref{sec:res}).
%

Looking in more detail, let us first consider the two bottom panels
of Fig.~\ref{fig:fdelnew}, which show all available high-quality
data for $^{12}$C and the data with momentum transfers $q>500$
MeV/c. We observe that many (but not all) of the data indeed tend to
collapse into a single function close to the $\Delta$ peak. The
spreading of the data is larger than what was observed in similar
analyses of QE data, but a tendency to cluster (scale) is seen, at
least for a subset of the data. In order to discuss scaling, and the
breaking of it, one may consider three different regions. First, in
the positive $\psi_\Delta'$ region the spreading of the data
increases and the data themselves tend to diverge. This behavior is
analogous to what happens for the QE case for large values of
$\psi'$ and it is due to the presence of contributions coming from
higher resonances. Second, for $\psi_\Delta'<-1$ the data form a
relatively uniform background showing no specific pattern. This is
the range where effects from 2p-2h MEC are expected to play a
significant role (see below). Finally there is the region
$-1<\psi_\Delta'<0$, where the spreading of the data is somehow less
evident and where both type of scale-breaking effects can
contribute.

In a first attempt to disentangle these effects, in the top
right-hand and two middle panels of the figure we apply a
progression of cuts on the data, specifically taking those with
$0.5$ GeV/c $< q < q_{cut}$, where $q_{cut}$ goes from 1 to 2 GeV/c.
As the cut tightens we expect to have fewer and fewer contributions
from higher inelasticities. For completeness, and for a better
understanding of the whole figure, in the upper left-hand panel we
also report the data for low momentum transfer ($q<0.5$ GeV/c).

The results shown in the different panels seem to indicate that the
presence of contributions from higher inelasticities corresponds to
values of $f^{non-QE}$ which lie  above the average scaling function
for $-0.5 < \psi_\Delta' < 0$ and below it for $-1 < \psi_\Delta' <
-0.5$ (for instance, compare the top and middle right-hand panels).
In particular we observe data sets which seem to cross the average
function around $\psi_\Delta'=-0.6$. They correspond to JLab cross
section data taken at an incident energy of 4.045 GeV and scattering
angles between 23 and 74 degrees, for which there is indeed  a
strong overlap of the $\Delta$ and higher inelastic contributions.

The observations above suggest that, if we are interested in
obtaining a phenomenological SuSA scaling function for the $\Delta$
region alone, $f^{non-QE}_{SuSA}$, these highly inelastic data sets
should be excluded from the fit. Such a fit, similar to that
obtained in~\cite{amaSSM04}, is indicated in Fig.~\ref{fig:fdelnew}
by the solid line, and in Fig.~\ref{fig:phenofits} it is compared
with the phenomenological fit for the QE region and, for reference,
with the RFG scaling function.

We observe that $f^{non-QE}_{SuSA}$ differs significantly from
$f^{SSM-QE}$. This is expected, because, besides incorporating
initial-state dynamics, the phenomenological non-QE scaling
functions certainly contain additional effects, such as those due to
the finite width of the $\Delta$ resonance, as well as potential
2p-2h MEC contributions. However it is interesting to investigate
whether or not these differences can be explained only in terms of
kinematics and of trivial effects, such as the finite width of the
$\Delta$, or whether also differences in the nuclear dynamics at the
QE and $\Delta$ peaks can contribute to them. In order to address
this issue we need to introduce some model for the cross sections in
the $\Delta$ region, and we will present this in the next section.

Let us conclude this section by introducing the
phenomenological SuSA model for the $\Delta$ region~\cite{amaSSM04}
mentioned in the introduction. Following the approach used in
previous section for the QE case, we can obtain the response
functions for $\Delta$ excitation from Eqs.~(\ref{eq:rltdel0}), by
substituting the RFG expression for $f^\Delta$ with the
phenomenological fit obtained from the data, namely \be
  R_{L,T}^{SuSA-\Delta,}(\kappa,\lambda)
  =\frac{1}{k_F}f^{non-QE}_{SuSA}(\psi_\Delta) G_{L,T}^{\Delta}\,.
  \label{eq:rltdelsusa}
\ee This model was tested in \cite{amaSSM04} for electron scattering
over a range of kinematics, showing agreement with the data at the
level of 10\% or better.

\section{SSM-based models for the inelastic region}
\label{sec:deltamodel}
In this section we develop a model for the response functions in the
inelastic region lying above the QE peak, basing the approach on the
assumption of universality of the superscaling function, {\it i.e.,}
using the same SSM approach employed for the QE region. This will
allow us to address two issues. On the one hand we will explore the
origin of the difference between the phenomenological scaling
functions obtained by fitting the data for the QE ($f^{SSM-QE}$) and
$\Delta$ ($f^{non-QE}_{SuSA}$) regions, as discussed in the last
section. We will start by assuming that this difference can be
accounted for only by kinematics and finite width effects, and we
will compare the scaling function obtained under this hypothesis
with the experimental one. On the other hand, we will investigate
further the role played by contributions from higher resonances in
producing the scaling violations shown by the experimental
$f^{non-QE}$ in the region $-1<\psi_\Delta'<0$.

As the model we present here is based on the impulse approximation,
it will not allow us to investigate MEC effects in the
$\psi_\Delta'<-1$ region directly, but it will turn out to be useful
later in Sec.~\ref{sec:res}, in presenting the experimental data in
a different and more focused way.

\subsection{Formalism}
\label{sec:forma2}

We follow closely the approach of \cite{bar03}, where a microscopic model based
on the RFG and on superscaling was used to study highly-inelastic
electron-nucleus scattering. The RFG expressions for the inelastic
nuclear response functions can be written as  \cite{bar03}:
\be
   R_{L,T}^{inel}= \frac{1}{k_F}\int_{\mu_{1}}^{\mu_{2}} d\mu_X \mu_X
   f_{RFG}(\psi_X) G_{L,T}^{inel}\;,
\label{eq:rltinel0}
\ee
where $\psi_X$ is obtained from Eqs.~(\ref{eq:psidelta}) and
(\ref{eq:rhodel})  for a generic invariant mass $W_X$ of the final
state reached by the nucleon, namely by replacing $\mu_\Delta$ with
$\mu_X = W_X / m_N$. The quantities $G_{L,T}^{inel}$, neglecting
terms of order $\eta_F^2$ and higher as before, are given by
\ba
G_L^{inel} &=& m_N \frac{\kappa}{2 \tau}\left\{
Z\left[(1+\tau\rho_X^2)\wtil_2^p - \wtil_1^p\right]
+ N \left[(1+\tau\rho_X^2)\wtil_2^n - \wtil_1^n\right]\right\}
\label{eq:glinel}\\
G_T^{inel} &=& m_N\frac{1}{\kappa}\left\{Z \wtil_1^p + N \wtil_1^n\right\}\,,
\label{eq:gtinel}
\ea
where $\tilde{w}_{1,2}$ are the inelastic single-nucleon structure
functions, which depend on two variables, the four-momentum transfer
$Q^2$ and the invariant mass $W_X$ or, equivalently the
single-nucleon Bjorken variable $ x=|Q^2|/[W_X^2-m_N^2-Q^2]$ (see
also \cite{bar03}). Note that the inelastic structure functions have
dimension of $E^{-1}$, at variance with the previous QE and $\Delta$
cases: for this reason we indicate them as $\tilde{w}$. The
integration limits in Eq.~(\ref{eq:rltinel0}) are given by \ba
\mu_1 &=& 1 + \mu_\pi \nonumber\\
\mu_2 &=&  1 +2 \lambda -\epsilon_S \label{eq:intlim} \ea with
$\mu_\pi= m_\pi / m_N$ and where $\epsilon_S=E_S/m_N$ is the
dimensionless version of the nucleon separation energy. The first
limit is simply the threshold for pion production, while the second
was derived in~\cite{bar03}.

Following a procedure analogous to that illustrated for QE
scattering, we can now generalize the RFG by making the substitution
\be
f_{RFG}(\psi_X) \rightarrow f^{SSM-QE}(\psi_X)\equiv f^{SSM-non-QE}(\psi_X)\equiv f^{SSM}(\psi_X)
\ee
in Eq.~(\ref{eq:rltinel0}). This modeling, which we will call
SSM-inel in the following, is thus based on the
assumption,  suggested by the RFG,  that there exists only {\em a single
universal scaling function} and that the latter can be identified
with the phenomenological fit obtained from the QE longitudinal data. Henceforth,
for simplicity we denote the phenomenological (super-) universal scaling function
to be used both for impulsive QE and inelastic contributions by $f^{SSM}$.

Important ingredients of the model are, of course, the
single-nucleon structure functions. In our past work~\cite{bar03},
which was focused on the highly-inelastic scattering region, we used
the Bodek {\it et al.}~\cite{bodek} parametrizations of the proton
and neutron structure functions which were available at the time.
However, in recent years new studies, both theoretical
\cite{Benhar:2006nr,lalapas} and experimental
\cite{BC,BC1,Bosted:2007hw,Psaker:2008ju}, of the nucleon structure
functions in the resonance region have been performed, indicating
the need for more sophisticated parametrizations. As we are now
studying this region,  we have updated our calculations using more
modern expressions for $\tilde{w}_{1,2}^{p,n}$. We thus use
parametrizations recently obtained by Bosted and Christy
\cite{bostedPC}, both for the proton \cite{BC} and neutron
\cite{BC1} structure functions. We note that all details regarding
the nucleon resonances, such as finite widths, are automatically
included in the parametrization and that we do not consider any
possible medium-modification of single-hadron properties.

\begin{figure}[th]
\includegraphics[scale=0.75,  bb= 100 100 500 750]{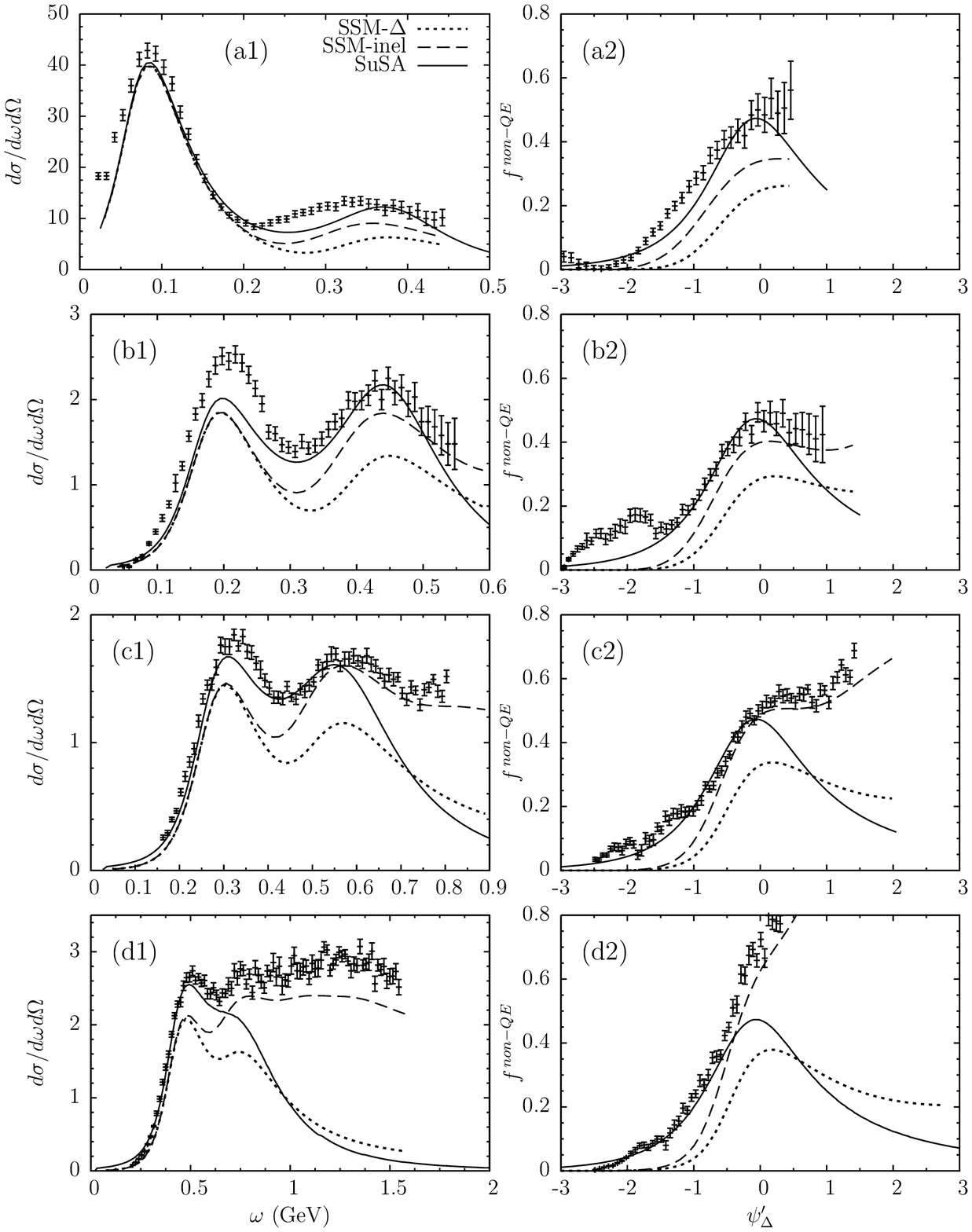}
\caption{Cross sections in nb/sr/MeV (left-hand panels; SSM-QE
results also included) and non-QE superscaling
  functions  (right-hand panels) for $^{12}$C, calculated within
  the SSM-inel (dashed line) and \ssmdg (dotted) approaches, and compared
with the phenomenological SuSA results (full line). The kinematics
selected here are summarized in Table I; the data are taken from
from~\cite{Barreau:1983,Sealock:1989,Day:1993}. } \label{fig:xsect}
\end{figure}

In order to understand better the role played by higher resonances, we
also consider a variant of the full SSM-inel model. For this approach, denoted
SSM-$\Delta$, we consider contributions coming only from $N\rightarrow\Delta$
excitations, which we describe in terms of form factors, and we take the
finite width of the $\Delta$ explicitly into account.
Following~\cite{ama99} we start with
\be
    R_{L,T}^{\Delta} =
    \int_{\mu_1}^{\mu_2} \frac{1}{\pi}
    \frac{\Gamma(\mu_X)/2 m_N}
    {\left(\mu_X-\mu_\Delta\right)^2 + \Gamma(\mu_X)^2/4 m_N^2}
    R_{L,T}^{\Delta}(\kappa,\lambda,\mu_X)d \mu_X\,,
\label{eq:ssmdg}
\ee
  where $R_{L,T}^{\Delta}(\kappa,\lambda,\mu_X)$ are the
  RFG response functions of Eq.~(\ref{eq:rltdel0})
  calculated using  a generic nucleon excitation invariant mass
  $\mu_X$, and 
  $\psi_X$
  is obtained from Eq.~(\ref{eq:psidelta}) for
  $\mu_\Delta \rightarrow \mu_X$.
  Once again we then generalize the RFG model by substituting
  for the RFG scaling function in Eq.~(\ref{eq:rltdel0}) the universal one,
  $f^\Delta(\psi_X) \rightarrow f^{SSM}(\psi_X)$.
The integration limits in Eq.~(\ref{eq:ssmdg})
  are those of Eqs.~(\ref{eq:intlim}), and
  the $\mu_X$ dependence of the $\Delta$ width $\Gamma$ is given by

\be
     \Gamma(\mu_X) = \Gamma_0 \frac{\mu_\Delta}{\mu_X}
     \left(\frac{p^\star_\pi}{p^{res}_\pi}\right)^3
\ee
  with $\Gamma_0=120$ MeV,
\be
   p^\star_\pi = \frac{m_N}{\mu_X}\left[
   \frac{(\mu_X^2 - 1 - \mu_\pi^2)^2}{4} -  \mu_\pi^2
   \right]^{\frac{1}{2}}
   \label{eq:pstar}
\ee

and where $p^{res}_\pi$ is obtained from Eq.~(\ref{eq:pstar})
with $\mu_X=\mu_\Delta$. We then compute inclusive cross sections
using the response function in Eq.~(\ref{eq:ssmdg}) and, in order to
obtain superscaling functions within this model, namely
$f^{SSM-\Delta}(\psi_\Delta)$, as usual we divide the cross sections
by $S^{\Delta}/k_F$, where $S^{\Delta}$ is the factor given in
Eq.~(\ref{eq:sdelta}). These superscaling functions may then be
compared with the phenomenological SuSA one,
$f^{SSM-inel}_{SuSA}(\psi_\Delta)$, discussed above
[Eq.~(\ref{eq:fdelta})].

\subsection{Results}
\label{sec:results}
In this section we illustrate the results for the superscaling
function obtained using the SSM-inel and SSM-$\Delta$ models,
together with the phenomenological SuSA fit. Before studying the
behavior of the non-QE scaling function over the whole range of
kinematics considered in Fig.~\ref{fig:fdelnew}, we will present a
few selected examples of cross sections and scaling functions. The
use of cross sections allows a direct comparison with ``real'' and
more familiar data, and the selection of fixed kinematics can
illustrate better the characteristics, and the limits, of the
models. This comparison is shown in Fig.~\ref{fig:xsect}, where the
left-hand panels show results for cross sections and the right-hand
panels show results for non-QE scaling functions. For illustration,
we choose to consider kinematics covering a limited range of energy
and momentum transfer, large enough so that  the $\Delta$ excitation
is clearly present and small enough so that higher inelastic
contributions do not overlap completely with the QE and $\Delta$
peaks. Specifically, we select a lower limit case (panels a1, a2)
corresponding to incident energy $\epsilon=620$ MeV and scattering
angle $\theta=36^\circ$, and an upper limit case (panels d1, d2)
with $\epsilon=3595$ MeV and $\theta=16^\circ$. In order to explore
the angle dependence of the cross sections and scaling functions, in
the middle panels we show results for intermediate kinematics with
two choices of scattering angle, $\epsilon=680$ MeV,
$\theta=60^\circ$ (panels b1,b2) and $\epsilon=1299$ MeV,
$\theta=37.5^\circ$ (panels c1,c2). The kinematics are summarized in
Table I which contains as well the momentum transfers at the QE and
$\Delta$ peaks, $q_{QE}$ and $q_\Delta$, respectively.

\begin{table}[tbp] \centering%
\begin{tabular}{|l||l|l|l|l|}
\hline Case & $\epsilon $ [MeV] & $\theta $ [deg] & $q_{QE}$ [MeV/c]
& $q_{\Delta }$ [MeV/c] \\ \hline\hline
a & 620 & 36 & 366 & 460 \\ \hline
b & 680 & 60 & 606 & 600 \\ \hline
c & 1299 & 37.5 & 791 & 850 \\\hline
d & 3595 & 16.02 & 1056 & 1189 \\ \hline
\end{tabular}%
\caption{$^{12}$C$(e,e')$ kinematics considered}\label{TableI}%
\end{table}%

In the figure we compare the results obtained using our SSM-inel and
SSM-$\Delta$ models with those corresponding to the SuSA fit introduced at
the end of Sec.~\ref{sec:scaldel}. The SSM-$\Delta$ model is certainly an
overly simple one and, as can be seen from Fig.~\ref{fig:xsect},
the corresponding curves show the largest discrepancies with the
data. However, the \ssmdg results are qualitatively interesting
because, when compared with the SSM-inel results, they allow us to
some extent to disentangle the effects related to contributions
arising from higher resonances, which cannot be eliminated from the
data.

The cross sections plotted in the left-hand column of the figure
include the QE contribution calculated within the SSM-QE modeling
outlined in Sec.~\ref{sec:scalQE}, which is the same for all models.
Differences between the various curves in the QE region are
therefore due to differences in the non-QE part of the cross
sections obtained using the various models. By looking at the  cross
sections we can clearly see that our inelastic model always
underestimates the data in both QE and, especially, inelastic
regions. More specifically, for small incident energy (upper panels)
the QE peak is well reproduced. At the $\Delta$ peak both SSM models
clearly underestimate the data, while SuSA obviously reproduces the
peak reasonably, since it was fit to the data. All models are unable
to reproduce the cross section completely in the region between the
QE and $\Delta$ peaks.

Similar results hold for the $\Delta$-peak region at larger
scattering angles (panel b1). We notice that in this case the SSM
and SuSA modeling underestimates the data even at the QE peak. This
is related to the fact that at large scattering angles the
transverse contribution is dominant. Previous scaling
studies~\cite{DS199,DS299} in fact showed that the transverse QE
superscaling function extracted from the data differs from the
longitudinal one and exhibits stronger scaling violations. We
attribute these differences to contributions beyond the impulse
approximation, such as 2p-2h MEC and correlations, which are not
included in the models discussed in this section (see, however,
Sec.~\ref{sec:res}). Moreover, at larger angles the overlap between
the QE and $\Delta$ peaks becomes more significant, which explains
the difference between SuSA and SSM models at the QE peak.

The same considerations can be extended to the case of higher
incident energies (panels c1 and d1). We observe that in these cases
the SSM-$\Delta$ results decrease very rapidly at large energy
transfer, as does the SuSA curve, because no higher inelastic
contributions beyond the $\Delta$ are included. The SSM-inel curve
has an $\omega$-dependence similar to that of the data for large
energy transfer, suggesting that the single-nucleon inelastic
content has been correctly implemented in the model. However, the
experimental cross sections are again underestimated even in the
higher inelastic region.

With these considerations about cross sections in mind, we can now
examine the right-hand panels of the figure, which show the non-QE
scaling functions. We can summarize our findings as follows. As
already said, the SSM-inel model always underestimates the data.
This difference, in both size and shape, is particularly relevant
for small incident energies and, at all kinematics, for relatively
large negative values of the scaling variable $\psi_\Delta'$. At
very low energy (upper panels of the figure) or for
$\psi_\Delta'<-1$ this is expected, because in these regions effects
stemming from correlations and 2p-2h MEC can play an important
role~\cite{ADP03,ADP04} and they cannot be reproduced by models
which assume impulsive, quasi-free scattering on bound nucleons.

In the region $-1<\psi_\Delta'<0$ the theoretical SSM-inel curves
still fall below the data, but their shape is similar to that
displayed by the experimental scaling function. The discrepancies
are larger below $\psi_\Delta'=-0.5$, where 2p-2h MEC may still
contribute sizably, whereas when approaching the $\Delta$ peak the
theoretical curves lie closer to the data. The conclusion we draw
from these observations is that the basic idea of the
phenomenological superscaling-based model (SSM-inel) is probably
correct and that it can account for most of the difference in shape
between the experimental QE and non-QE scaling functions, but that
the model presented here is still too simple and needs some
improvements in order to be considered quantitatively reliable. In
particular, as previously observed, the model assumes universality
of the longitudinal and transverse QE scaling function, {\it i.e.,}
the so-called scaling of the 0th kind, which has been shown to be
violated by the QE data. While part of this violation can be
ascribed to correlation and 2p-2h MEC effects, as discussed in the next section,
a certain amount of it could be present even at the impulse
approximation level, and, if so, should be incorporated in the model
by using different scaling functions for the T and L responses. This
would lead to a renormalization of the calculated cross sections and
non-QE scaling functions, which may fill some of the discrepancy
with the data at the $\Delta$ peak. Unfortunately, such an
improvement of the model is not straightforward, although work is
now in progress along this line.

\begin{figure}[ht]
\includegraphics[scale=0.75,  bb= 100 150 500 800]{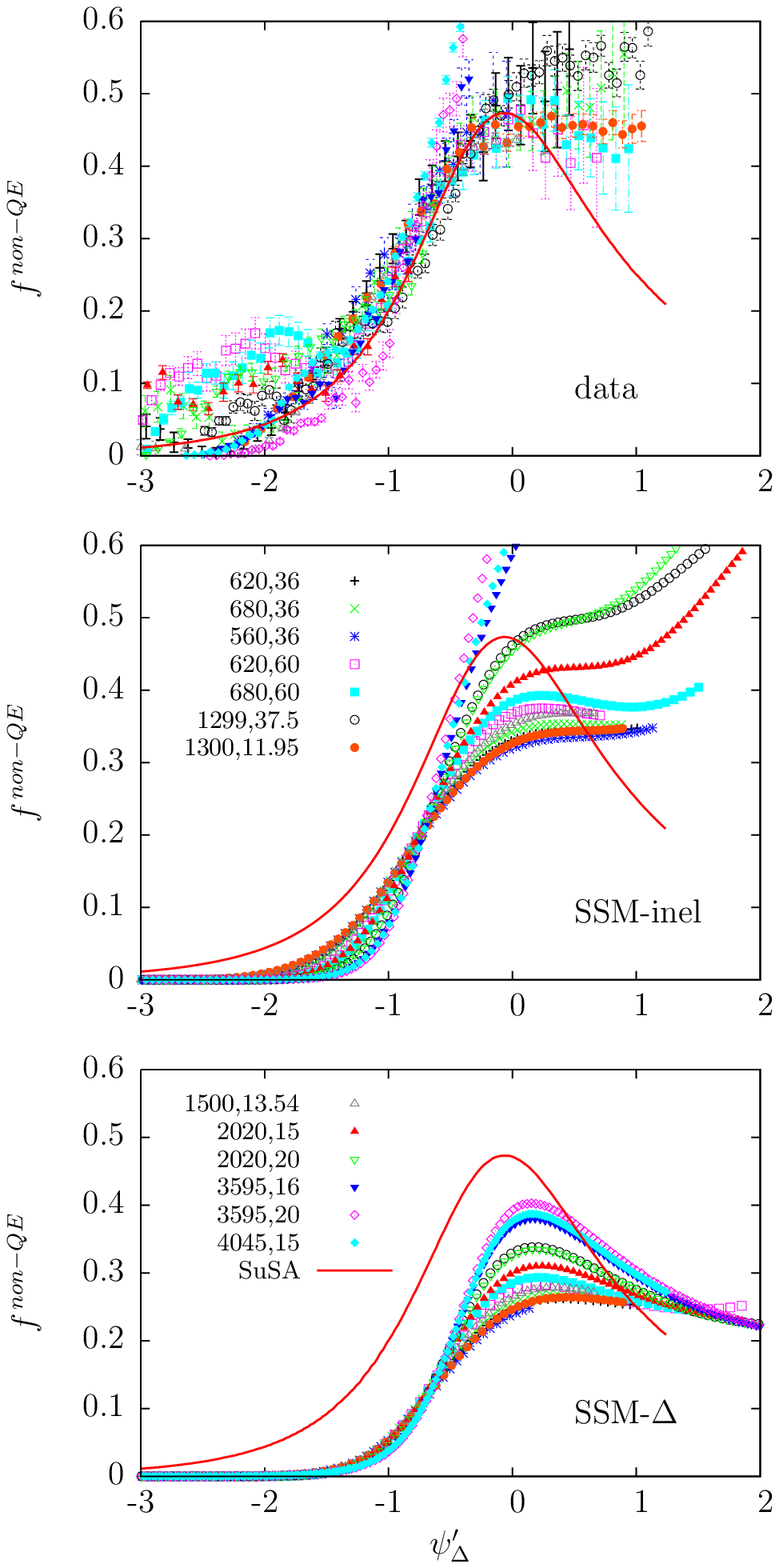}
\caption{(Color online) Experimental ``QE subtracted'' data for
$f^{non-QE}$ for $^{12}$C for a variety of kinematics (top  panel)
and corresponding results of the SSM-inel model (middle) and of  the
\ssmdg model (bottom). The kinematics considered are labeled with
$\epsilon$ (MeV) and $\theta$ (deg). The SuSA fit is shown by the
solid (red) line. The values of the momentum transfer
for the kinematics presented here fall approximately in the interval
$0.5<q<1.5$ GeV/c.} \label{fig:fdelsumm}
\end{figure}

If we accept that at least some of the difference in normalization
between the data and the calculated $f^{non-QE}$ close to the
$\Delta$ peak can be accounted for by an improvement in the scaling
functions used as ingredients in the model, then the SSM-inel
results obtained so far can provide some useful additional insight on the
behavior of the $\Delta$ superscaling function. In
Fig.~\ref{fig:fdelsumm} we plot the function $f^{non-QE}$ for a
relatively large set of kinematics (indicated in the key inside the
figure) corresponding approximately to values of the momentum
transfer in the range 500-1500 MeV/c. The top panel shows the
experimental non-QE scaling functions, the middle panel those
obtained within the SSM-inel model and the bottom panel those
calculated with the \ssmdg model. We see that both the data and the
SSM-inel scaling functions present the same type and degree of
scaling violations in the region $-1<\psi_\Delta'<0$, with the
curves corresponding to the highest momentum transfer being the
lowest ones for approximately $\psi_\Delta<-0.5$ and then becoming
the highest one for larger values of the scaling variable. In contrast, this
behavior is practically absent in the \ssmdg results,
suggesting that scaling violations in the region $-1<\psi_\Delta'<0$
are essentially due to contributions from higher resonances. This
observation has important consequences for SuSA
modeling of neutrino cross sections~\cite{amaSSM04}, because it
supports the validity of using the universal scaling function $f^{SSM}$
in predicting cross sections for kinematical conditions
in which only contributions up to the excitation of the $\Delta$
resonance are relevant.

Still looking at Fig.~\ref{fig:fdelsumm}, let us mention that both
SSM models provide scaling-violations at the $\Delta$ peak which
seem to be larger than those exhibited by the data. In our
study we have checked that this is due to kinematical
effects, being related to the interplay between integration limits
and the dependence of the variable $\psi_X'$ upon the invariant mass
$\mu_X$. The inclusion of some degree of scaling-violation in the
phenomenological  scaling function used in the model may solve this
problem.
\begin{figure}[th]
\includegraphics[scale=0.75,  bb= 100 300 500 750]{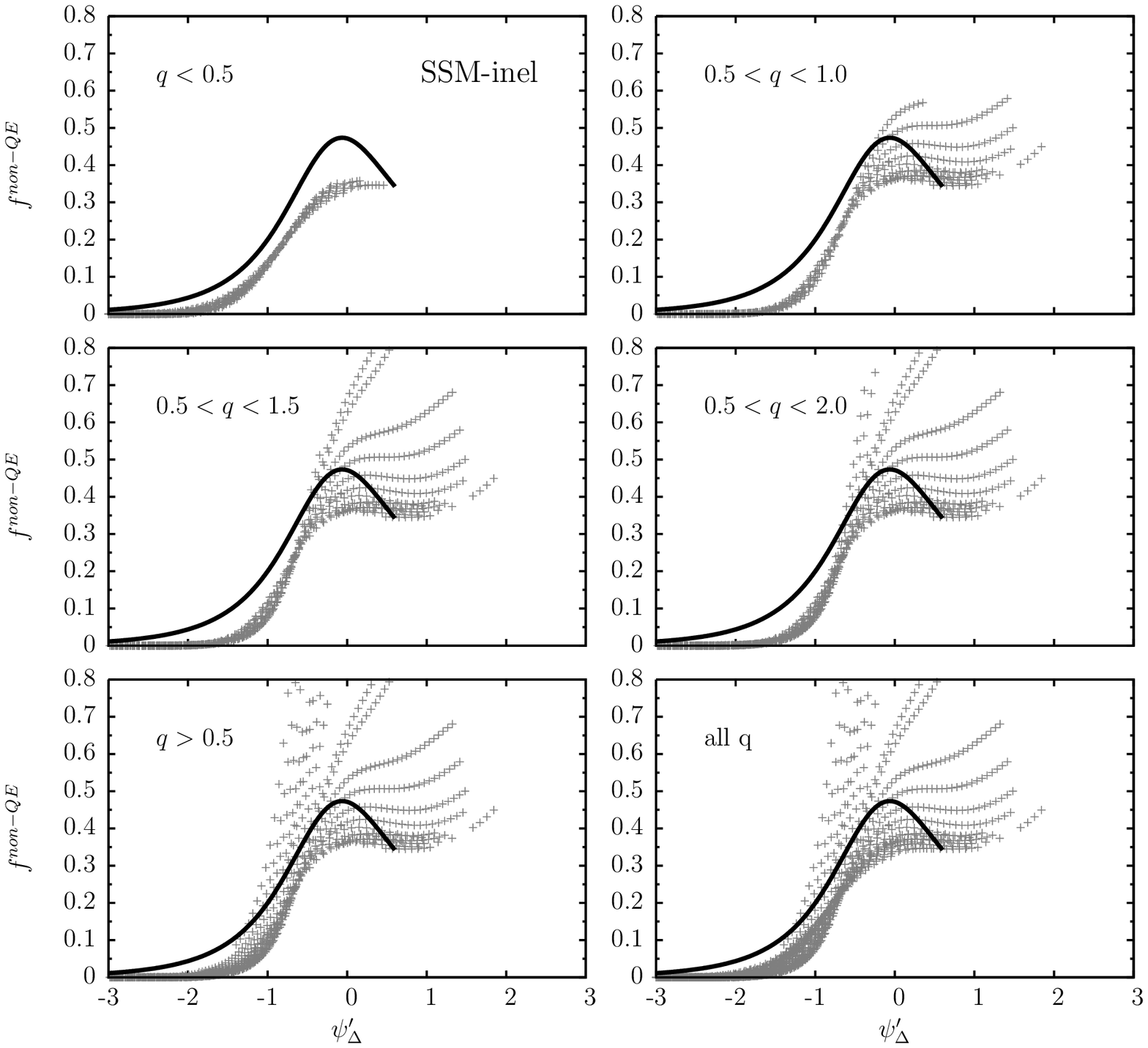}
\caption{Non-QE superscaling
  function for $^{12}$C calculated within the SSM-inel model
versus the $\Delta$ scaling variable $\psi_\Delta'$.  The same
kinematical cuts as in Fig.~\ref{fig:fdelnew} are considered, as
indicated in the different panels. The phenomenological fit of the
non-QE superscaling function is also shown for comparison (solid
line).  } \label{fig:finelmod}
\end{figure}
%

The differences in the behavior of the theoretical and experimental
scaling functions at the peak of the $\Delta$ may also be related to
the different role of final-state interactions (FSI) for QE
scattering and $\Delta$ excitation. In fact, previous studies in the
QE region have shown that the phenomenological QE scaling function
is affected by FSI at the right of the QE peak where FSI produce a
tail, and partially at the peak, since a larger tail at positive
$\psi$ results in a smaller maximum value of the scaling function.
While the tail of the phenomenological function contributes very
little to the calculated non-QE scaling function at the left of the
$\Delta$ peak due  to the limits of integration (see
Eqs.~(\ref{eq:rltinel0}, \ref{eq:intlim} and \ref{eq:ssmdg})), its
maximum value may have some relevance at the $\Delta$ peak. Some
details concerning the limits of integration and the role of
 $f^{SSM}$ in determining the non-QE scaling
function can be found in the Appendix.
%
\begin{figure}[th]
\includegraphics[scale=0.75,  bb= 100 300 500 750]{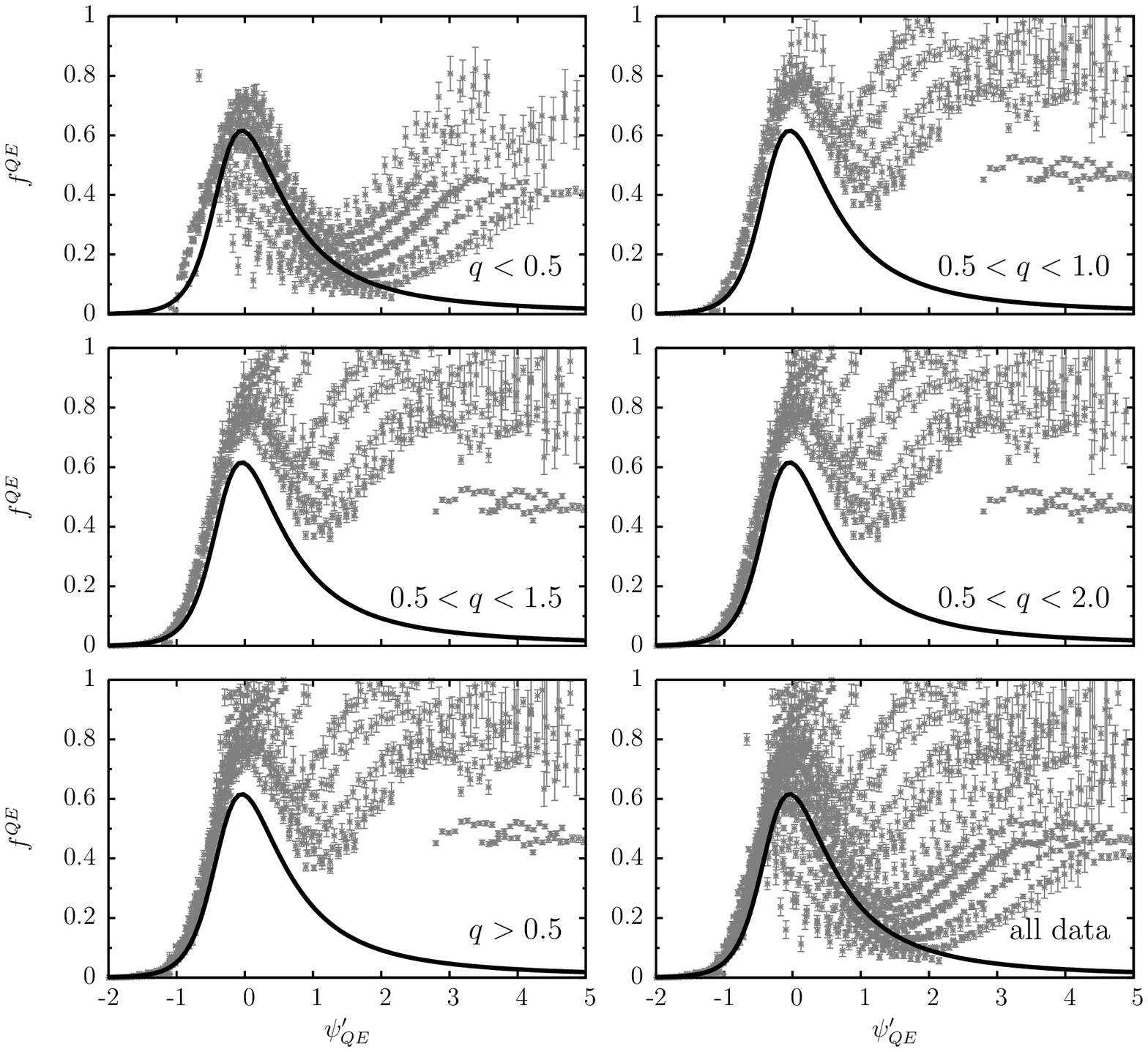}
\caption{Full quasielastic scaling function $f^{QE}$
 for $^{12}C$ as a function
of $\psi_{QE}'$.  The same kinematical cuts as in
Fig.~\ref{fig:fdelnew} are considered, as indicated in the different
panels. The solid line is the phenomenological QE fit used in this
work. } \label{fig:fqe}
\end{figure}

Finally, before proceeding in the next section to the analysis of
the residual after impulsive contributions have been removed, and to
complete the overview of our results for the function $f^{non-QE}$,
we show in Fig.~\ref{fig:finelmod} the complete set of SSM-inel
results for all kinematics where data are available, with the same
kinematical cuts used for the results presented in
Fig.~\ref{fig:fdelnew} (see also Table I). Also, for comparison, in
Fig.~\ref{fig:fqe} we show $f^{QE}$ (Eq.~(\ref{eq:fqenew})) as a
function of $\psi_{QE}'$.

\section{Residual non-impulsive contributions and synthesis of the cross section}
\label{sec:res}

\begin{figure}[t]
\includegraphics[scale=0.8,  bb= 130 440 500 750]{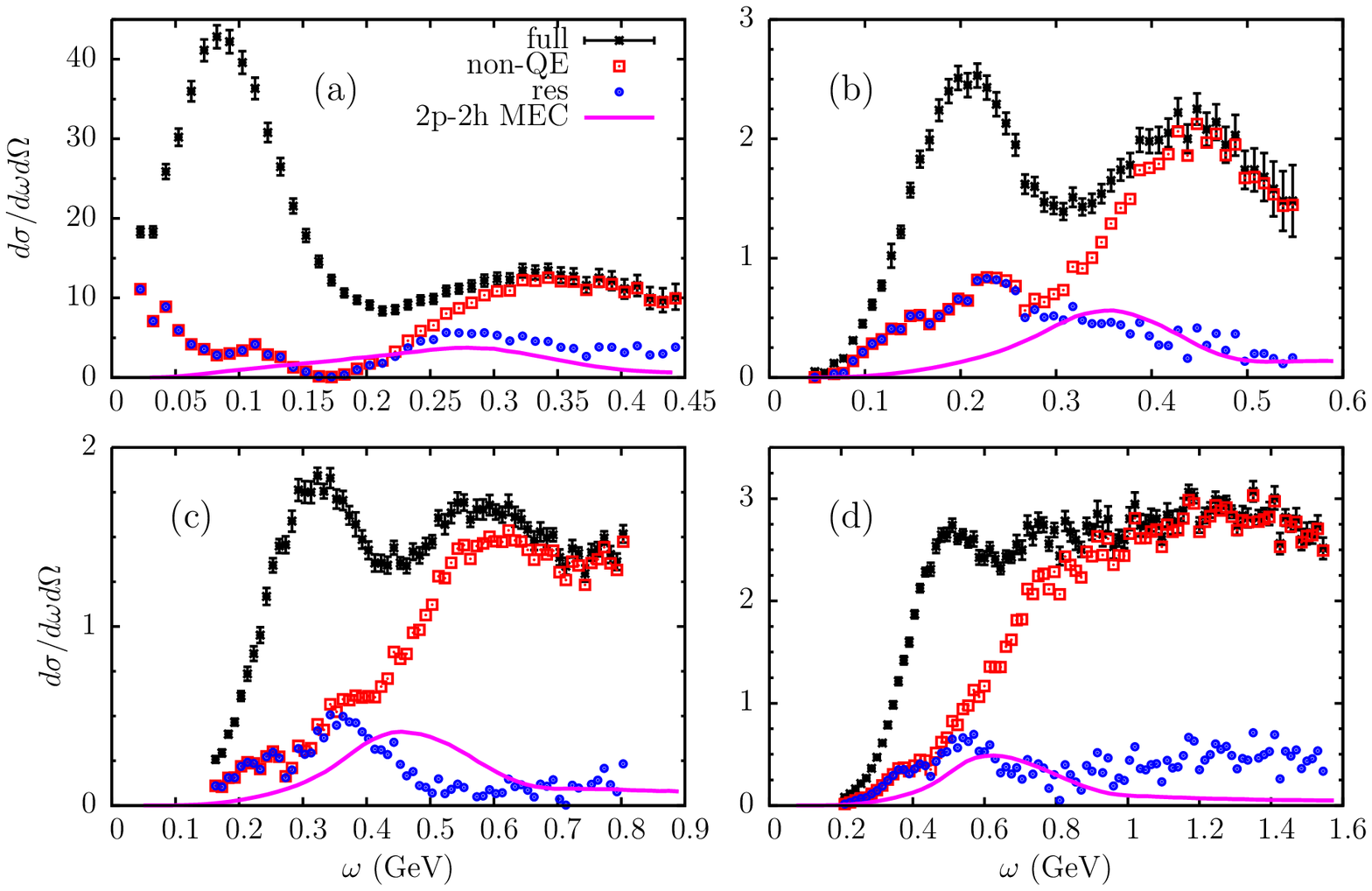}
\caption{(Color online) Cross sections in nb/sr/MeV  versus $\omega$. The
kinematics corresponding to the various labels are the same as in
Fig.~\ref{fig:xsect} and Table I. The curves are discussed in the
text. } \label{fig:xsectres}
\end{figure}

The superscaling-based model developed in the previous sections
allows one to study the behavior of the superscaling function within
the context of the impulse approximation and therefore to assess the size
of any potential non-impulsive contributions. In particular, it is
interesting to combine the two impulsive contributions denoted
SSM-QE (Sec.~\ref{sec:scalQE}) and SSM-inel (Sec.~\ref{sec:forma2})
and subtract this from the data to yield a residual:
\be \left( \displaystyle{ \frac{d\sigma}{d\epsilon'd\Omega} }
\right)^{res} \equiv \left( \displaystyle{
\frac{d\sigma}{d\epsilon'd\Omega} } \right)^{exp} - \left(
\displaystyle{ \frac{d\sigma}{d\epsilon'd\Omega} } \right)^{SSM-QE}
- \left( \displaystyle{ \frac{d\sigma}{d\epsilon'd\Omega} }
\right)^{SSM-inel}  \,. \label{eq:sigres} \ee
The results are shown in Fig.~\ref{fig:xsectres} for the same
kinematics considered in Fig.~\ref{fig:xsect}.  Here the (black)
stars are the complete experimental data, the (red) squares the
QE-subtracted cross sections [Eq.~(\ref{eq:sigdelequiv})] and the
(blue) circles the residual cross sections [Eq.~(\ref{eq:sigres})].
Note that, lacking any means of evaluating what errors are incurred
in the subtraction procedures, we have not given any uncertainties
for the non-QE and residual cross sections shown in the figure.

Focusing on the residual cross sections we see significant
contributions left over after the SSM-QE and SSM-inel results have
been removed. As stated several times in the previous sections we
expect there to be non-impulsive effects from 2p-2h
MEC~\cite{ADP03,ADP04}. Indeed, when these are compared with the
residuals (shown as solid magenta curves in the figure) one sees
rough agreement. That is not to say that one now has a fully
satisfactory picture of inclusive electron scattering in this
kinematic region --- there are still several open issues. In
particular, when MEC effects are included (and they are not
optional; they must be included) gauge invariance requires that
corresponding correlation contributions must also occur.
In~\cite{albe89,ama01, Amaro:2002mj,Amaro:2003yd,ama09} this problem
was dealt with for the 1p-1h sector. However, this has not yet been
done for the 2p-2h response, although work is in
progress~\cite{ADPfuture} to address this issue. Another issue goes
back to comments made in the previous sections, namely, even the
SSM-QE approach has some uncertainties in that scaling of the 0th
kind may be broken to a small degree, and that the somewhat larger
transverse scaling functions found in the RMF approach and the 1p-1h
MEC contributions (which lead to a small reduction of the transverse
cross section) may not completely compensate one another. In effect
one could break the 0th-kind scaling by using a slightly different
scaling function for the transverse contributions and thereby modify
the residuals seen in the figure. It is clear, however, that a
significant amount of the residual can be explained by the 2p-2h MEC
contributions. In the last section we shall return to this point and
comment on the implications this has for predicting neutrino
reaction cross sections.

In this section we focus primarily on the cross sections and only at
the end of the section we will briefly return to discuss the non-QE
scaling function in order to assess the validity of SuSA-based
models. Note that we should not expect the 2p-2h MEC contributions
to scale using either type of scaling discussed above, {\it i.e.,}
either the QE type or the $\Delta$ type. In fact these contributions
have their own characteristic scaling behavior and work in progress
is aimed at exploring this behavior in the residual data. With these
comments in mind, let us work in the opposite direction and, rather
than analyzing the cross section, attempt to synthesize it using the
three types of contributions.

\begin{figure}[t]
\includegraphics[scale=0.8,  bb= 130 440 500 720]{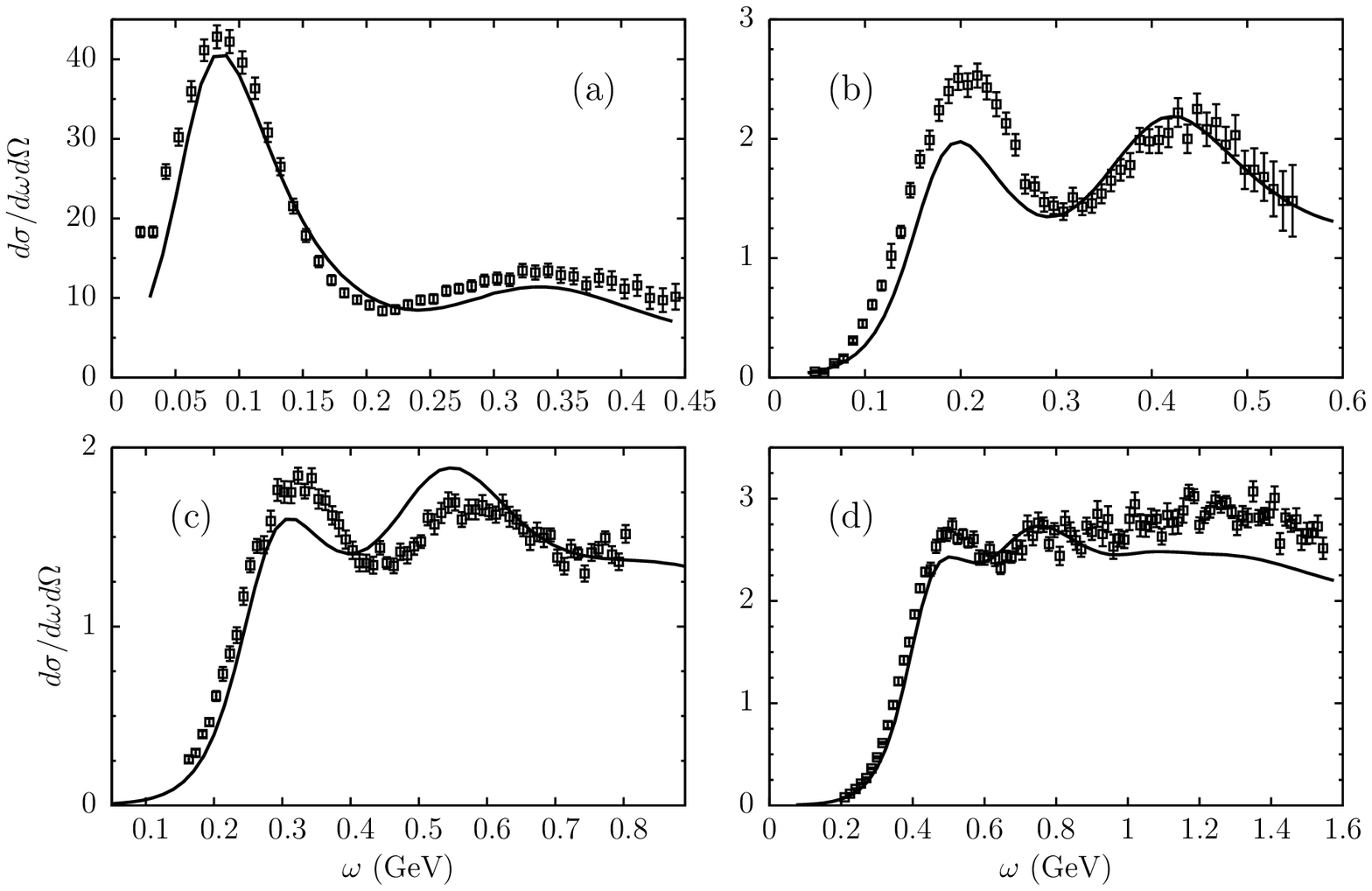}
\caption{Cross sections in nb/sr/MeV  versus $\omega$. The
kinematics corresponding to the various labels are the same as in
Fig.~\ref{fig:xsect} and Table I. The curves are the sum of the
SSM-QE, SSM-inel and 2p-2h MEC contributions. }
\label{fig:syn-total}
\end{figure}

In Fig.~\ref{fig:syn-total} we show the net result of adding
together the SSM-QE, the SSM-inel and the 2p-2h MEC contributions
for comparison with the data. The results are quite encouraging: the
basic qualitative structure of the data is also present in the net
result of the superscaling analysis, although clearly there is more
to be done before one can claim to have a fully quantitative
description of inclusive electron scattering in this region of
kinematics. In particular, the net result of adding the three
contributions falls short of the data in the QE peak region, and
this might be fixed by slightly breaking the 0th-kind scaling (as
discussed above) or by exploiting the flexibility that is inevitably
present in the modeling of the 2p-2h MEC contributions (for
instance, by using a different shift energy than the one that was
chosen for the results presented here). It should be stressed that
this rather good level of agreement between theory and experiment
has been obtained by adding together three separate contributions,
each with its own distinctive kinematic dependence, and thus any
attempt to represent experimental data using only a subset of the
contributions is bound to fail for some choice of kinematics.

\begin{figure}[t]
\includegraphics[scale=0.9,  bb= 130 520 500 750]{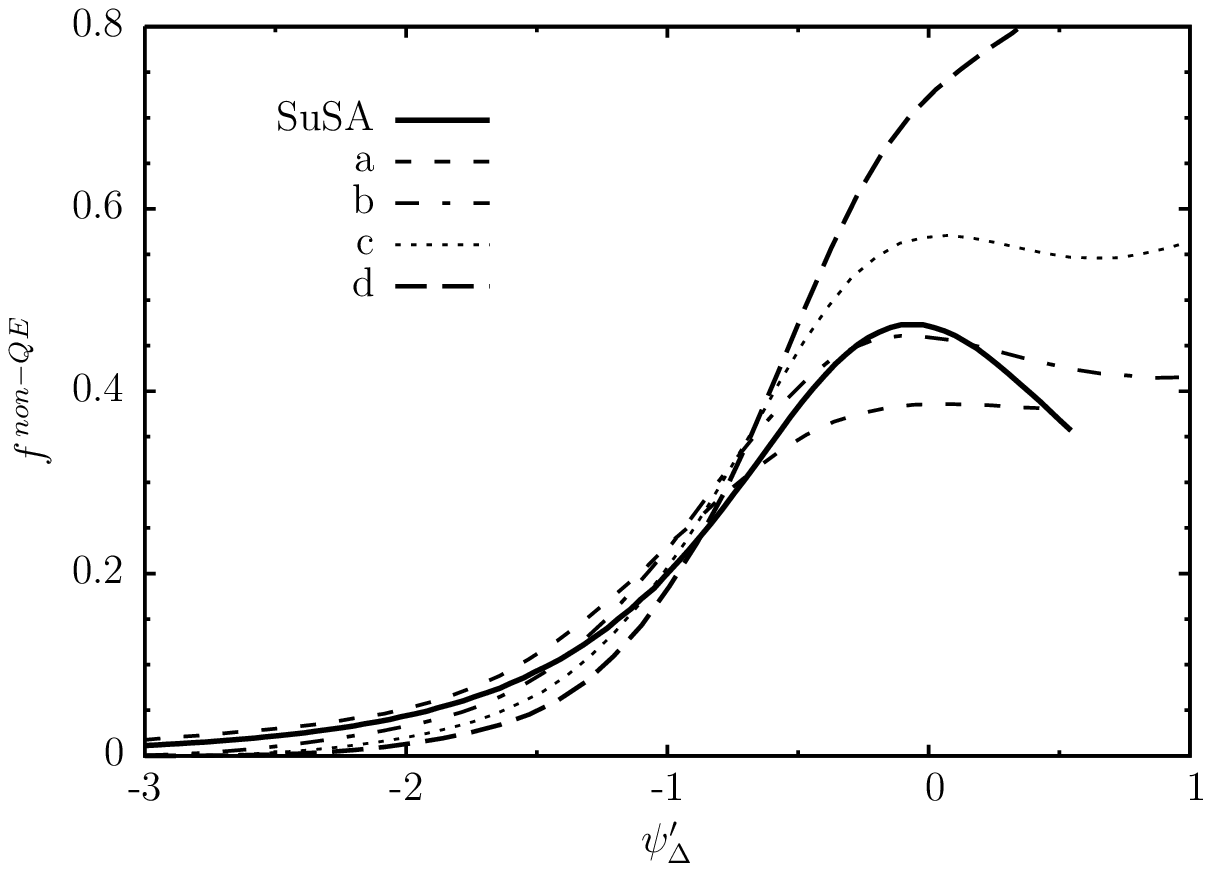}
\caption{Non-QE superscaling functions calculated by using
the sum of SSM-inel and  2p-2h MEC contributions to the cross sections
in Eq.~(\ref{eq:fdelta}). Labels a, b, c, d correspond to the
kinematics used in Fig.~\ref{fig:xsect} and listed in Table~\ref{TableI}.
The solid line is the SuSA fit.}
\label{fig:fig10}
\end{figure}

In order to make contact with the Superscaling Analysis of
\cite{amaSSM04}, we conclude this section by taking the non-QE
superscaling functions obtained by using the sum of SSM-inel and
2p-2h MEC cross sections and inserting them in
Eq.~(\ref{eq:fdelta}). These are shown in Fig.~\ref{fig:fig10}, for
the kinematics considered in the previous figures (see
Table~\ref{TableI}). The point of doing this, despite the concluding
statements made in the preceding paragraph, is to provide a
comparison with the phenomenological SuSA results discussed in
Sec.~\ref{sec:scaldel}. We observe that the inclusion of 2p-2h MEC
contributions brings the calculated non-QE scaling function closer
to the phenomenological fit, supporting the validity of the
SuSA-based model for lepton-nucleus cross sections at kinematics
dominated by $\Delta$ excitation. However, the strength shown by the
residual data close to the QE peak, not accounted for by the
theoretical MEC curves considered in this section (as discussed
above), affects the non-QE scaling functions at $\psi'_\Delta$
values below approximately -1.5. This can be seen by examining the
right-hand column of Fig.~\ref{fig:xsect}, for instance. These
effects should be carefully considered in the future when
constructing quantitatively reliable models.

\section{Summary and conclusions}
\label{sec:conclu}

In this work we have explored superscaling in electron-nucleus
scattering. We have started by reviewing the procedures for
analyzing scaling in the quasielastic region. A universal
longitudinal QE scaling function emerges, both based on
phenomenology and on modeling. Upon assuming that 0th-kind scaling
is satisfied (universality of transverse and longitudinal scaling
functions) we arrive at our model, denoted SSM-QE, for these
contributions.

Next we have focused on the region lying to the right of the QE
peak, first introducing the definition of the experimental scaling
function in this region, $f^{non-QE}$. This entails subtracting the
SSM-QE scaling predictions from the data. We have studied the
scaling behavior of $f^{non-QE}$ by analyzing all available
high-quality data for $^{12}$C. We have found reasonable scaling
below the $\Delta$ peak, with scaling violations that can be mainly
explained in terms of contributions coming from higher resonances.
Following the SuSA approach presented in \cite{amaSSM04}, we have
obtained a phenomenological fit, $f^{non-QE}_{SuSA}$, which differs
from the phenomenological function $f^{SSM-QE}$ obtained in previous
studies of QE scattering.

In order to understand this difference, and to explore in detail the
breaking of scaling shown by $f^{non-QE}$, we have developed an
extension for inelastic electron-nucleus scattering within the
impulse approximation denoted $f^{SSM-inel}$ which is based on
previous studies of the same type~\cite{bar03}. The model begins
with the formulation of the response functions in the RFG model, and
extends the latter by incorporating in them the universal scaling
function $f^{SSM}$ obtained from fits of QE scattering data, making
the approach in a sense super-universal. The entire inelastic
response on the nucleon is incorporated using a recent
representation of the nucleon's structure functions for kinematics
going from pion-production threshold to where DIS takes over
\cite{bostedPC}, both for the proton \cite{BC} and neutron
\cite{BC1}.

The comparison of this impulsive model with the experimental data, for both
cross sections and non-QE scaling functions, is good, but not
entirely satisfactory at first glance, since the results always fall
below the data. However, the acceptable agreement of the shape of
the calculated scaling function with the data, suggests that the
differences between the experimental scaling functions obtained in
the QE and $\Delta$ regions could be mainly explained in terms of
the kinematical effects discussed in the Appendix. Additionally, the results of the SSM-inel model
allow us to conclude that the scaling violations observed for
$-1<\psi_\Delta'<0$ can be mostly explained by the presence of
contributions from higher resonances. In particular, by comparing results from the full
SSM-inel model, in which the entire inelastic responses of the nucleons are included, with a
variant of this approach (denoted SSM-$\Delta$), in which only the $\Delta$ is included, it has been
possible to gain some insight into the roles played by excitations lying above the $\Delta$.

Having explored the superscaling properties of the SSM-QE/SSM-inel
model, we have used this model to subtract from the experimental
data both the QE contributions and those inelastic contributions
that can be described within the impulse approximation, thus
isolating non-impulsive contributions. When this residual is
compared with the known non-impulsive contributions, namely, those
arising from 2p-2h MEC, one sees improved agreement between modeling
and data. Indeed, it appears that the 2p-2h MEC contributions are
essential if one is to have a quantitative picture of inclusive
electron scattering at the kinematics considered in this work.

Finally, a few words are in order concerning the implications the
present study has for predictions of neutrino reaction cross
sections. Clearly all of the ingredients discussed here (QE,
inelastic and MEC contributions) also enter in studying the latter,
insofar as the vector current is concerned. The axial-vector current
required for neutrino reactions is another matter, however. Unlike
the polar-vector current where the leading-order MEC effects enter
as transverse effects, but not as longitudinal effects, the
axial-vector current is the opposite (due to the extra $\gamma_5$ in
the basic current, which switches the contributions of the upper and
lower components in the required matrix elements). Accordingly, for
the axial-vector currents there are no leading-order transverse
effects from MEC, while there are for the axial longitudinal/charge
currents. The latter are small for neutrino reactions at the
kinematics of interest in this work and consequently for neutrino
reactions the MEC effects enter asymmetrically --- essentially, only
via the polar-vector currents, but not the axial-vector currents. We
have seen that the (vector) MEC effects are significant and thus any
model that does not have them runs the risk of incurring errors of
typically 10--20\% in predicting neutrino cross sections. Using
overly simple models such as the RFG is adequate for crude estimates
of the neutrino cross sections, although almost certainly the SSM
analyses presented here are considerably better as they capture much
of the correct kinematical dependences of the polar-vector parts of
the electroweak nuclear response. However, as we have seen in the
present work these impulsive superscaling models do not entirely
capture all of the necessary content in the currents since they are
missing the non-impulsive MEC contributions. To the degree that the
latter are important one has a complicated problem containing at
least three parts, the SSM-QE and SSM-inel impulsive contributions
together with the 2p-2h MEC effects, each with its own distinctive
kinematic dependences.

\acknowledgments We are grateful to P. Bosted for providing the code
containing the new nucleon structure function parametrizations and to A. De Pace
for providing the 2p-2h MEC results shown in Sec.~\ref{sec:res}.
This work was partially supported by DGI (Spain):
FIS2008-01143, FPA2006-13807-C02-01, FIS2008-04189, by the Junta de
Andaluc\'{\i}a, by the INFN-MEC collaboration agreement
(project ``Study of relativistic dynamics in neutrino and
electron scattering'') and by the Spanish Consolider-Ingenio 2000
program CPAN (CSD2007-00042). It was also
  supported in part (TWD) by the U.S. Department of Energy under
  contract No. DE-FG02-94ER40818.

\appendix*
\section{The $\Delta$ superscaling function at the $\Delta$ peak}
In Sec.~\ref{sec:results} we observed that at the $\Delta$ peak the
scaling violations shown by the superscaling function obtained
within the SSM-inel model seem to be larger than those present in
the data. Comparable scale-breaking effects are also obtained within
the SSM-$\Delta$ model, which includes only the excitation of the
$\Delta$ resonance, and therefore they cannot be explained in terms
of an incorrect treatment of higher resonances. Here we show that
the origin of these scaling violations in our models is related to
kinematical effects and to the shape and value of the
phenomenological QE function used as input at its peak. In order to
do so, we work within the SSM-$\Delta$ model, whose simplicity
allows us to explore in detail the effects of the various terms
entering the formulae for the response functions and of the
corresponding integration limits.

Let us return to the lower panel of Fig.~\ref{fig:fdelsumm}, where
we observe excellent scaling for $\psi'_\Delta\le -0.5$, whereas a
significant amount of scaling violation remains at larger
$\psi'_\Delta$ and it increases approaching the peak.

\begin{figure}[t]
\includegraphics[scale=0.75,  bb= 100 250 500 800]{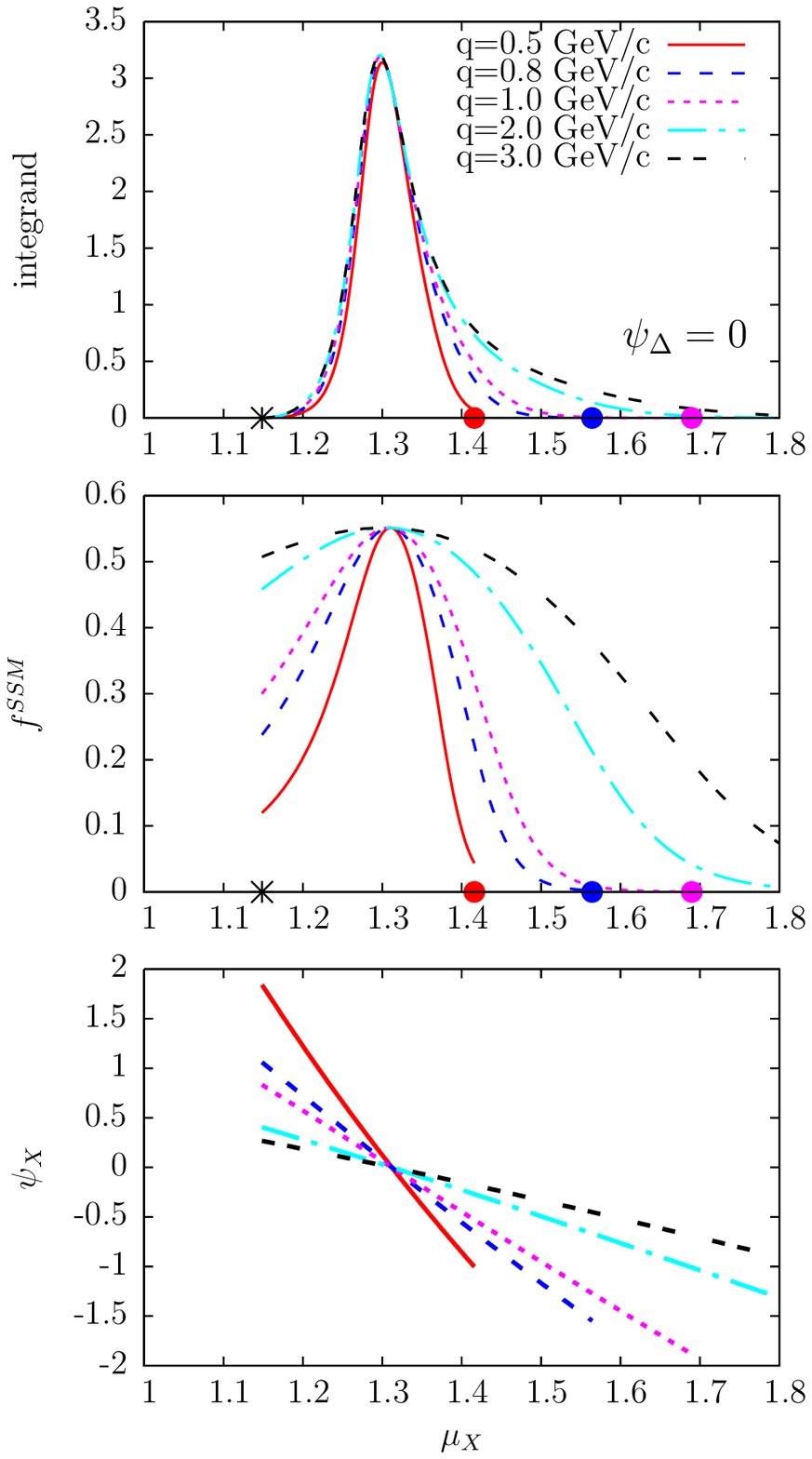}
\caption{(Color online) Upper panel: integrand of
Eq.~(\ref{eq:integral}), {\it i.e.,} of $f^\Delta_{appx}$, as a
function of $\mu_X$, for various values of the momentum transfer, as
indicated by the labels. The asterisk and points on the x-axis
indicate the lower and upper limits of integration, respectively.
Middle panel: function $f^{SSM}$ as a function of $\mu_X$. Lower
panel: inelastic scaling variable $\psi_X$ as a function of $\mu_X$.
All curves are calculated at fixed $\psi_\Delta=0$ (here, for
simplicity, $E_{shift}$ has been taken to be zero). }
\label{fig:int0}
\end{figure}

\begin{figure}[t]
\includegraphics[scale=0.75,  bb= 100 250 500 800]{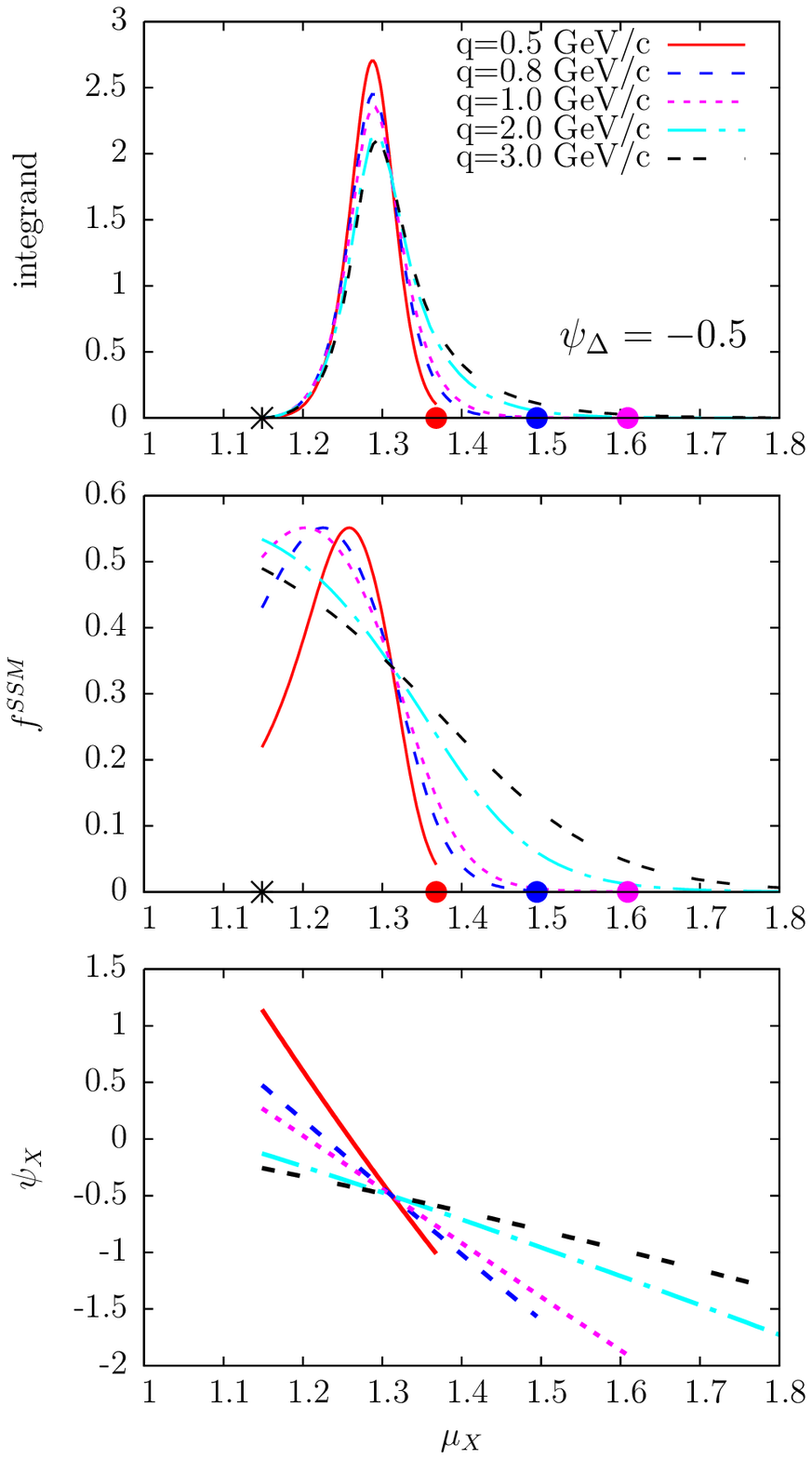}
\caption{(Color online) As for Fig.~\ref{fig:int0}, but now at fixed
$\psi_\Delta=-0.5$.}\label{fig:intm05}
\end{figure}

In order to understand this behavior let us examine the integral in
Eq.~(\ref{eq:ssmdg}) together with the formulae presented in
Sec.~\ref{sec:scaldel}. We can easily see that, except for minor
effects due to a residual $\mu_X$ dependence in the coefficients
multiplying the form factors in
Eqs.~(\ref{eq:w1del},\ref{eq:w2del}), the SSM-$\Delta$ superscaling
function $f^\Delta$ is essentially given by the integral
\be f^\Delta_{appx}(\psi_\Delta) \equiv
    \int_{\mu_1}^{\mu_2} \frac{1}{\pi}
    \frac{\Gamma(\mu_X)/2 m_N}{\left(\mu_X-\mu_\Delta\right)^2 +
      \Gamma(\mu_X)^2
/4 m_N^2}
    f^{SSM}(\psi_X) d \mu_X\,,
\label{eq:integral}
\ee
whose calculation indeed produces curves that are close to the full
$f^\Delta$ and have the same scaling properties. We can thus use the
simpler expression in Eq.~(\ref{eq:integral}) to investigate further
the origin of the scaling behavior of  $f^\Delta$ and the scaling
violations it exhibits at the $\Delta$ peak. To do so we choose two
fixed values of $\psi_\Delta$ (for simplicity we also set
$E_{shift}=0$), namely $\psi_\Delta=0$ ({\it i.e.,} the Delta peak)
and $\psi_\Delta=-0.5$ and look at the behavior of the integrand of
Eq.~(\ref{eq:integral}). This integrand is displayed in the upper
panels of Figs.~\ref{fig:int0} and~\ref{fig:intm05}, for
$\psi_\Delta=0$ and $-0.5$, respectively, for different values of
the momentum transfer $q$, as indicated by the labels. The asterisk
close to origin of the x-axis indicates the integration limit
related to pion-threshold, while the different dots on the x-axis
indicate the upper limit of integration. For the largest values of
$q$ considered here the latter falls outside the plotted range of
$\mu_X$  and therefore the corresponding dots do not appear in the
figures.


In the middle panels of the same figures we show $f^{SSM}$ as a
function of $\mu_X$ and in the lower panels we show how $\psi_X$
varies with $\mu_X$ for fixed $\psi_\Delta$. Note that $\psi_X$
decreases for increasing $\mu_X$. By looking at Fig.~\ref{fig:int0}
we can see that as $q$ increases, the dependence of $\psi_X$ on
$\mu_X$ becomes weaker and thus $\psi_X$ stays closer to the fixed
value of $\psi_\Delta$, which in this case is 0. This means that for
larger $q$ the integrand receives contributions mostly from
$f^{SSM}(\psi_X)$ close to its peak, while for smaller $q$ a more
extended range of $\psi_X$ values contributes. For this reason the
integrand turns out to be larger for the highest values of $q$ and
this gives rise to the behavior observed in Fig.~\ref{fig:fdelsumm}.

The same type of behavior of $\psi_X (\mu_X)$ is observed also for
$\psi_\Delta=-0.5$ (lower panel of Fig.~\ref{fig:intm05}), implying
that for high $q$ the variable $\psi_X$ stays close to the negative
value -0.5, thus remaining in the region where $f^{SSM}(\psi_X)$ is
small. In this case, however, this is compensated by a larger (with
respect to what occurs for smaller values of $q$) integration
interval and thus the values of $f^\Delta_{appx}$ for different $q$
are very close to each other. Moving to even more negative values of
$\psi_\Delta$ one could see that the larger integration interval
occurring for high $q$ is no longer able to compensate for the
smaller values of $f^{SSM}(\psi_X)$ involved in the integration, so
that the larger $q$ becomes, the smaller is the value of $f^\Delta_{appx}$
obtained.

Similar considerations could be applied to the integral in
Eq.~(\ref{eq:rltinel0}) entering in the SSM-inel model, although in
this case a direct comparison between $f^\Delta$ and
$f^\Delta_{appx}$ cannot be made, because of the different
single-nucleon ingredients entering the cross section and the
dividing factor $S^\Delta$ of Eq.~(\ref{eq:sdelta}), which defines
$f^\Delta$. In particular in this case stronger scaling violations
appear even in the negative $\psi'_\Delta$-region, which, as
discussed in the paper, are related to the presence of higher
resonance contributions in the inelastic single-nucleon structure
functions.


\end{document}